\documentclass[aps,prstab,twocolumn,groupedaddress]{revtex4}
\usepackage{amsmath}
\usepackage{tikz}
\usepackage{lipsum}
\usepackage{subfig}
\usepackage{color}
\usepackage{newtxtext,newtxmath,amsmath}
\usepackage{natbib}

\begin{document}

\title{A Numerical Approach to Designing a Versatile Pepper-pot Mask for Emittance Measurement\footnote{Accepted by Nuclear Inst. and Methods in Physics Research, A for publication. DOI:10.1016/j.nima.2019.162485}}
	
\author{O.~Apsimon\footnote{Corresponding author, email: oznur.mete@manchester.ac.uk}, B.~Williamson and G.~Xia}
\address{The University of Manchester, Manchester M13 9PL, United Kingdom \\ also The Cockcroft Institute of Accelerator Science and Technology, Warrington WA4 4AD, United Kingdom}

\begin{abstract}
The pepper-pot method is a popular emittance measurement technique for high intensity beams at low energy such as those generated by photo-injectors. In this paper, the  beam dynamics in the space charge dominated regime and analytical design criteria for a mask-based emittance measurement (pepper-pot method) are revisited. A tracking code developed to test the performance of a pepper-pot setup is introduced. Examples of such testing are presented with particle distributions that were generated using PARMELA under different focusing conditions. These distributions were numerically tested against a series of mask geometries suggested by analytical criteria. The resulting fine-tuned geometries and beam dynamics features observed are presented.
\end{abstract}

\maketitle

\section{introduction}

The pepper-pot method is a well-known technique for phase space characterisation at low energies, before energy boosting, where the space charge force is significant for high brightness beams \cite{Zhang, Anderson}. It is widely used in radio-frequency photo-injectors which produce high brightness electron beams for light sources \cite{PSI, SPARC} and test facilities for other large scale applications \cite{UCLA, FNAL, CLIC}. Today, this method continues to be popular to measure the phase space of electron beams generated by conventional and advanced accelerators \cite{AWAKE, VELA, AlphaX, AWA}.  
 
\begin{figure}[htb!] 
\centering
\captionsetup{justification=justified}
\includegraphics[width=0.45\textwidth] {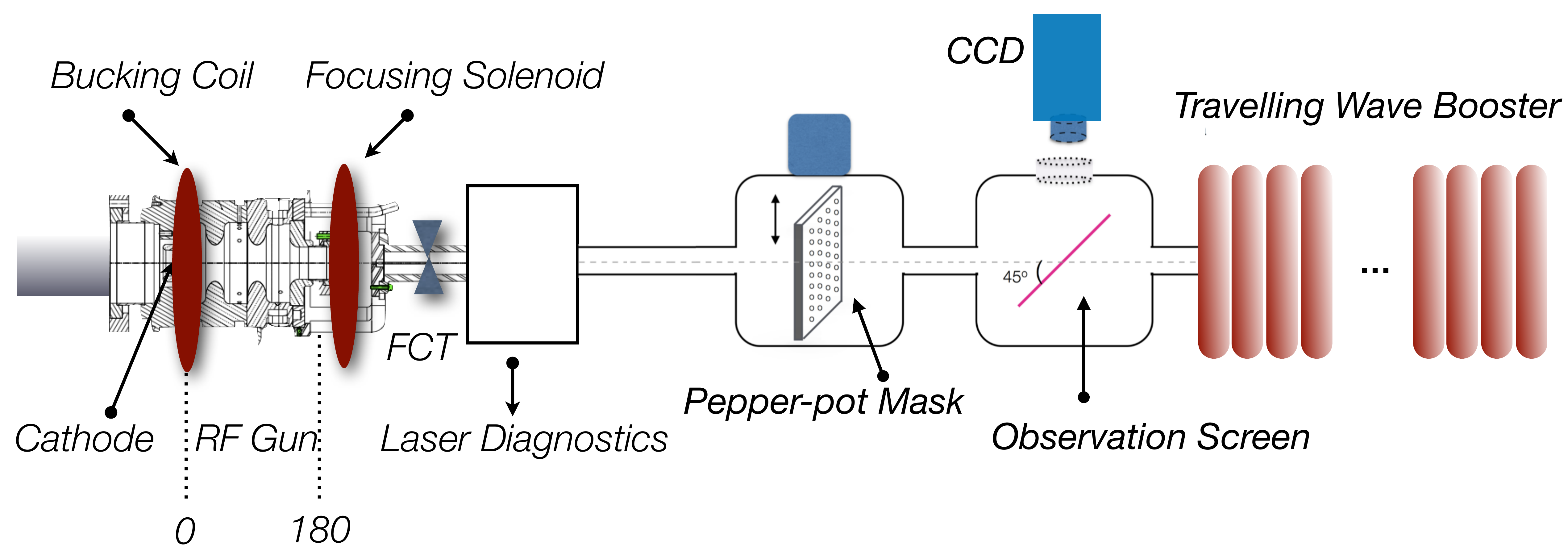}
\caption{The layout of a generic RF injector.}
\label{fig:layout}
\vspace{-1.5em}
\end{figure}

In this paper, we recapitulate the beam dynamics in the operating regime of the pepper-pot emittance measurement by studying the envelope equation for a space charge dominated beam, and numerically demonstrate the interplay between defocusing due to space charge and beam emittance. For these simulations, a typical photo-injector model was implemented in PARMELA \cite{PARMELA} as sketched in Fig. \ref{fig:layout}. This includes an RF gun at 3$\,$GHz working at an on-axis field of 100$\,$MV/m \cite{PHIN} followed by a travelling wave booster structure at the same frequency and working at 15$\,$MV/m \cite{Booster}. A pepper-pot setup is envisaged to be located after the RF gun prior to the booster structure (132$\,$cm after the cathode) and an analytical approach for its design is presented. It is discovered that a design purely based on analytical criteria does not always generate reliable results hence has to be validated and fine-tuned numerically. Consequently, a tracking algorithm is introduced to realise the transport through the mask and perform these validations. 

In terms of beam dynamics, the working point of a photo-injector generally is where beam is quasi-laminar at the start of the booster linac (in the vicinity of the mask) under appropriate focusing (provided by solenoids) which mainly performs the matching of the beam envelope. This ensures emittance compensation under space charge by aligning the phase space angle of each slice and minimising the projected emittance \cite{Serafini, Floettmann}. Both analytical and numerical design of a pepper-pot system are based on given beam parameters such as the rms beam size, divergence, emittance and energy. Therefore, naturally, the performance of pepper-pot measurements are optimised for this quasi-laminar regime. Under focusing conditions that does not fulfil emittance compensation, such as where the focal point of beam envelope occurs before or after the mask, (hence, an incoming beam is nonlaminar at the mask location) emittance measurements will include errors due to an underestimated beam divergence and beam size. Furthermore, the pepper-pot method analysis algorithm given in Section \ref{algorithm} assumes a Gaussian incident beam profile limiting the performance of the method to Gaussian, ideally quasi-laminar, beams. We tested various analytically suggested pepper-pot geometries using the above-mentioned numerical tool and investigated reconstructed emittance results under different solenoidal focusing in comparison to those retrieved from PARMELA simulations. As a result of these studies, we demonstrated that non-laminar beam envelopes or more specifically beams with un-compensated projected emittances might not be suitable for pepper-pot measurement.

\section{Beam Characteristics}
\label{characteristics}
In general, a photo-injector operates in a space charge dominated regime, generating a high intensity beam which is still at low energy (in the order of a few MeVs) before acceleration in a booster linac. For such a beam, the significance of space charge defocusing was semi-analytically studied by comparing it to the outward pressure of the beam associated with the normalised beam emittance. The envelope equation in paraxial limit is given as \cite{Serafini},
\begin{equation}
\sigma^{\prime\prime} + \sigma^{\prime}\frac{\gamma^{\prime}}{\beta^2\gamma}+K_r\sigma - \frac{\kappa_s}{\sigma\beta^3\gamma^3}-\frac{\varepsilon^2_n}{\sigma^3\beta^2\gamma^2} = 0,
\label{eqn:envelope}
\end{equation}
where $\sigma$ is the cylindrical symmetric rms beam size under effects of an external linear focusing channel with strength $K_r$, $\beta$ is the normalised mean beam velocity, $\gamma$ is the normalised beam energy in the units of the rest mass energy, $\kappa_s$ is the beam perveance and $\varepsilon_n$ is the normalised transverse emittance of a beam slice. In this equation, the last two terms represent the defocusing due to space charge and beam emittance, respectively. Here, beam perveance is given by $\kappa_s = I/2I_0$ where $I=Q/(2\sigma_z \sqrt{2ln(2)})$ is the peak beam current for a Gaussian beam and $I_0$ is a constant, known as Alfven current (17kA). The ratio of these two terms determines the dominant defocusing factor. The second and third terms represent the focusing due to adiabatic damping and external focusing.
\begin{figure}[htb!] 
\centering
\captionsetup{justification=justified}
\subfloat[]{\includegraphics[width=0.45\textwidth] {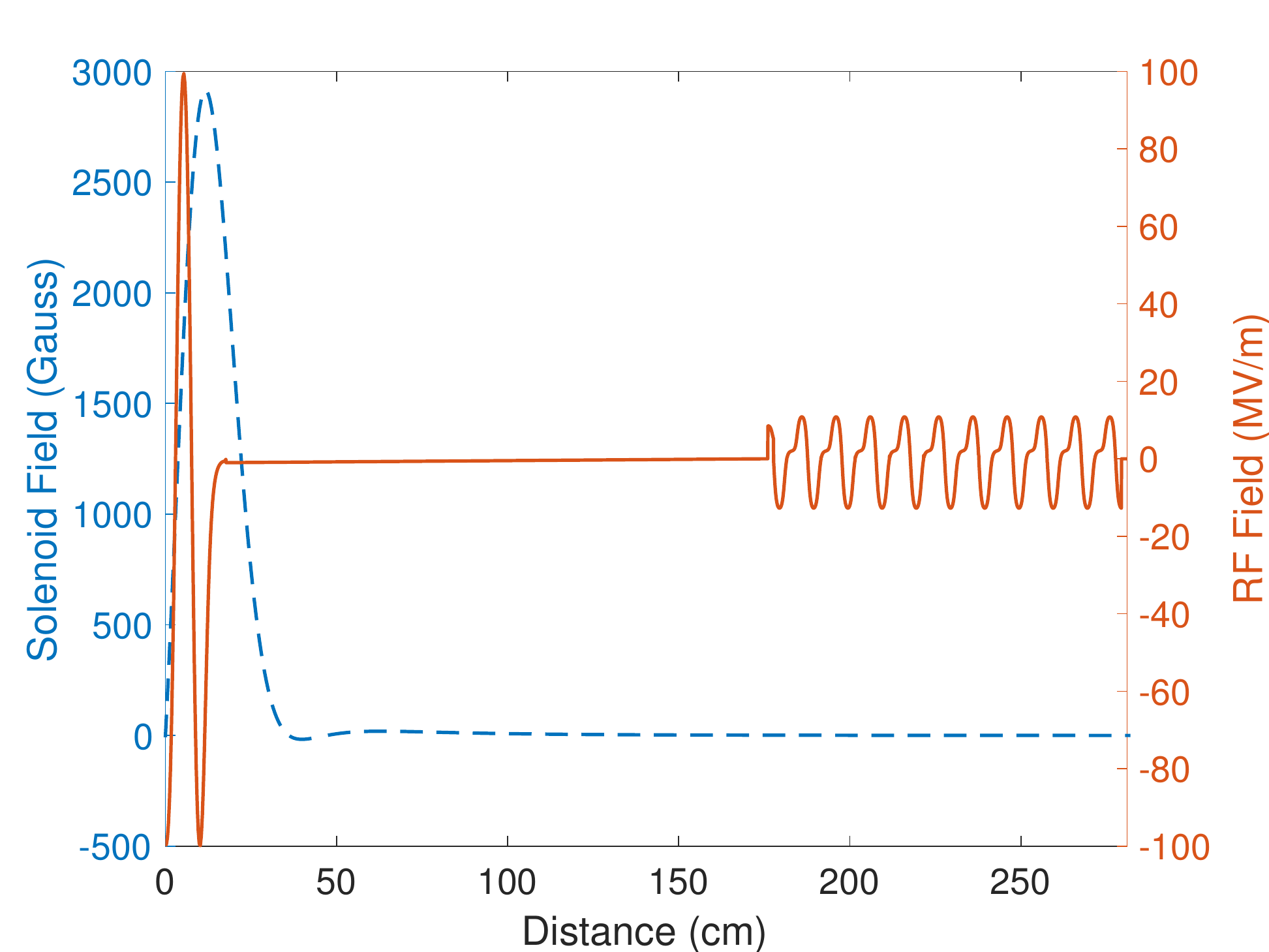}} \\
\subfloat[]{\includegraphics[width=0.42\textwidth] {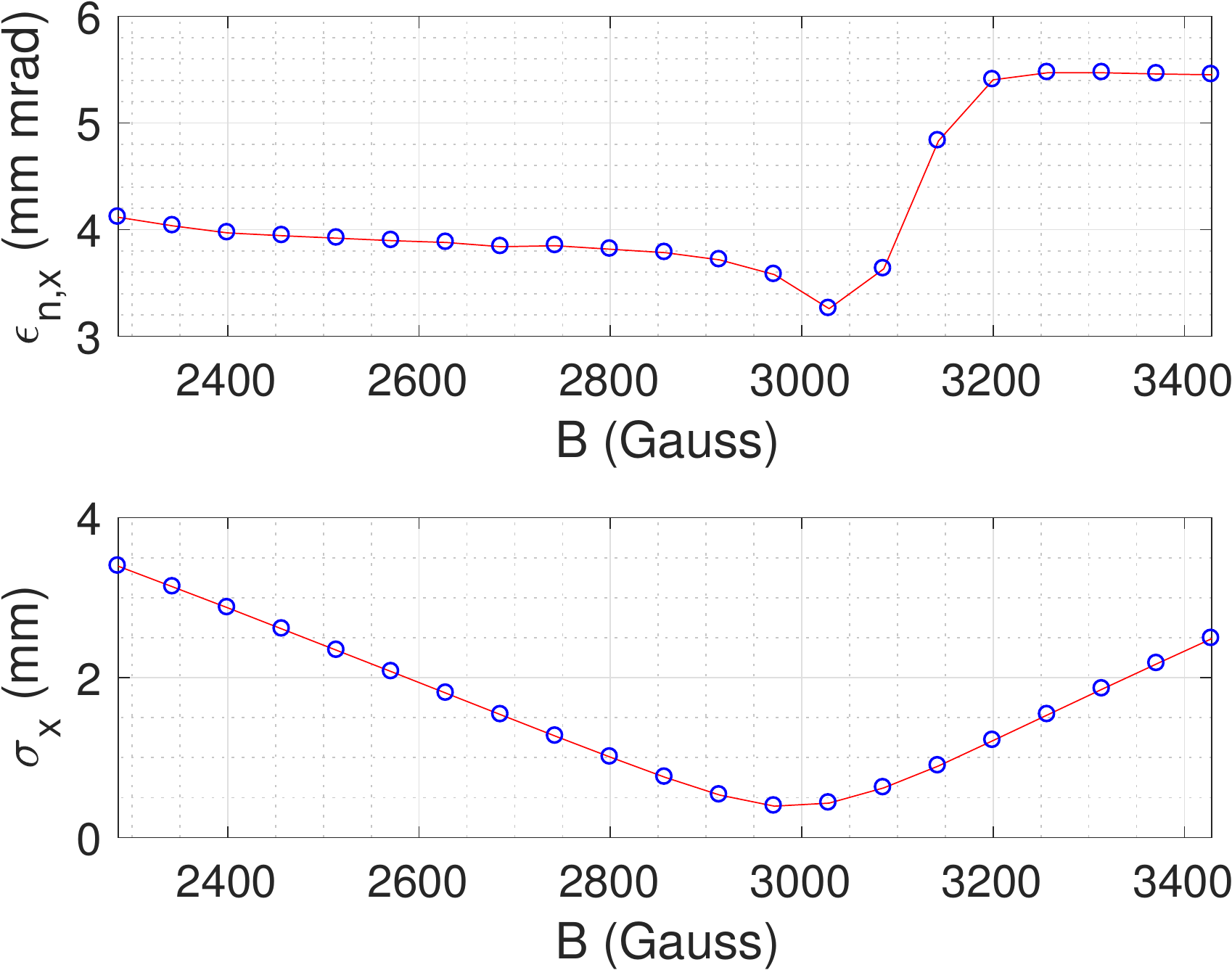}}  
\caption{(a) Electric field generated by the RF gun and booster linac at the on-crest phase (solid, blue curve) and magnetic field profile due to external solenoidal focusing (dashed, blue curve). (b) Normalised transverse beam emittance (top) and rms beam size (bottom) as a function of magnetic field.}
\label{fig:fields}
\end{figure}

The PARMELA generated electromagnetic (by the RF gun and booster linac) and magnetic fields (by the solenoids) acting on the beam for an envelope matched to the linac are presented in Fig. \ref{fig:fields}-a. The corresponding magnetic field profile in the figure has a peak value of approximately 2914$\,$G and this profile was scaled to achieve different amount of focusing around the RF gun. All results in this paper are presented for a 1$\,$nC beam reaching to an energy of 6.6$\,$MeV after the gun (and 17$\,$MeV after the linac which is not relevant at the mask location). Figure \ref{fig:fields}-b shows how beam emittance and rms size evolves as a function of solenoid current at the location of the mask. According to this, beam emittance is minimised at 3029$\,$G slightly after the beam focus at 2971$\,$G. Four working points were chosen across this focusing range; one that represents an under-focused envelope (2857$\,$G), a point with an envelope optimised to minimise the emittance at the location of the mask (3029$\,$G), another which ensures the matching to the booster linac (2914$\,$G) and finally a point with an over-focused envelope (3086$\,$G). These beam conditions are used to study the performance of different pepper-pot mask designs for different envelope characteristics which are summarised in Table \ref{tbl:characteristics}. 
\begin{table}[hbt!] 
\centering
\small
\caption{Characteristics of different beam envelopes that are simulated by PARMELA and used for the numerical design of a pepper-pot measurement system.} 
\resizebox{8cm}{!}
	{ 
   		\begin{tabular}{lccccc}
       		\hline
		Envelope type&$ B $ (Gauss) & $\sigma_x$ ($\mu$m) & $\sigma_x^{\prime}$  (mrad) & $\varepsilon_{n,x}$ (mm$\,$mrad) \\   
		\hline
       		\hline
		Under-focused		&	2857	&	754	&	0.4 	&	3.7\\
		Matched to Linac	&	2914	&	530	&	0.5	&	3.6\\
		Intermediate		&	2971	&	383	&	0.7	&	3.5\\
		Minimum at mask	&	3029	&	423	&	1.0	&	3.2\\
		Over-focused		&	3086	&	613	&	1.4	&	3.5\\
       		\hline		
   		\end{tabular}
   	}
\label{tbl:characteristics}
\end{table}

\begin{figure}[htb!] 
\centering
\captionsetup{justification=justified}
\subfloat[]{\includegraphics[width=0.4\textwidth] {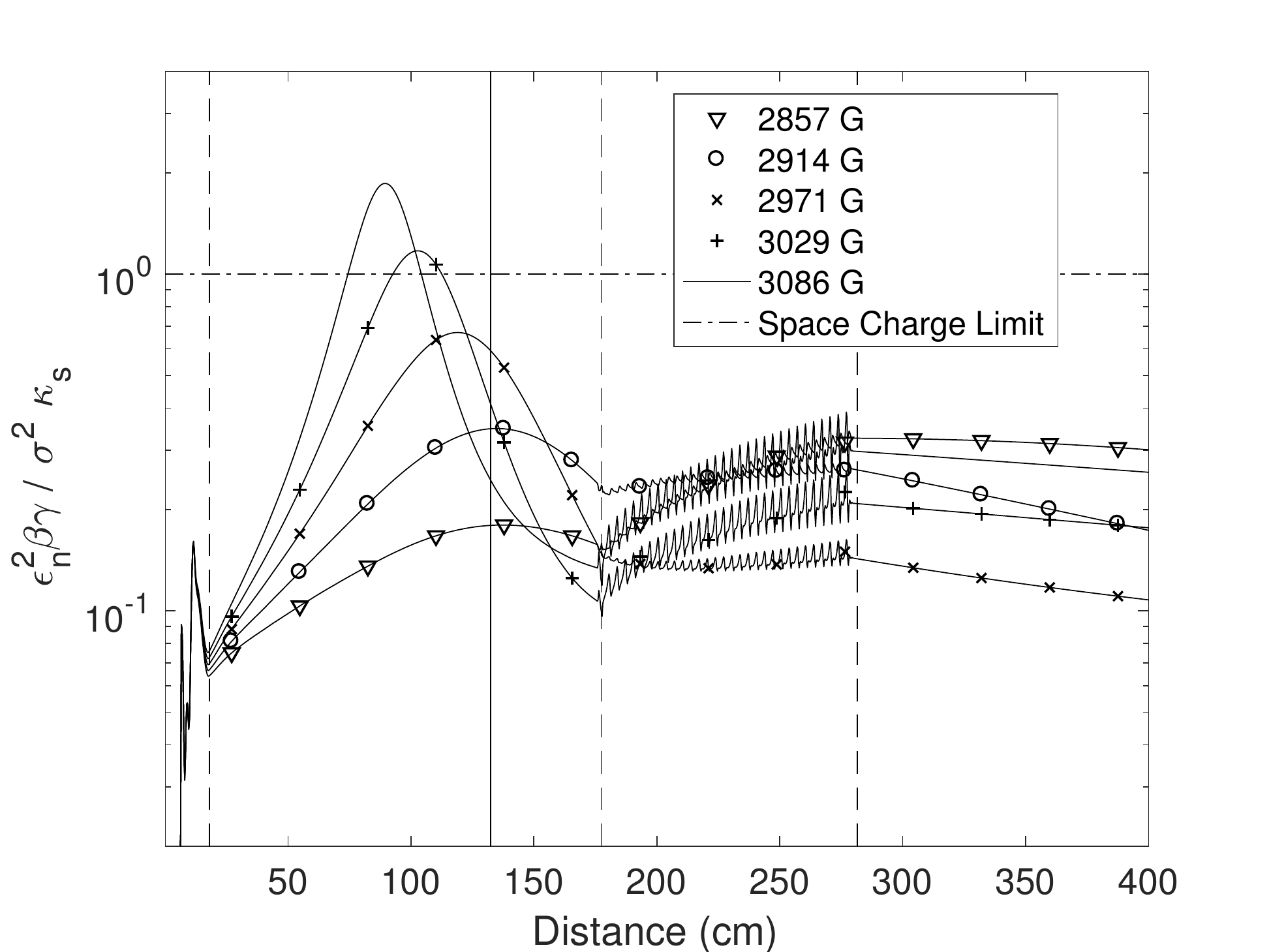}} \\
\subfloat[]{\includegraphics[width=0.4\textwidth] {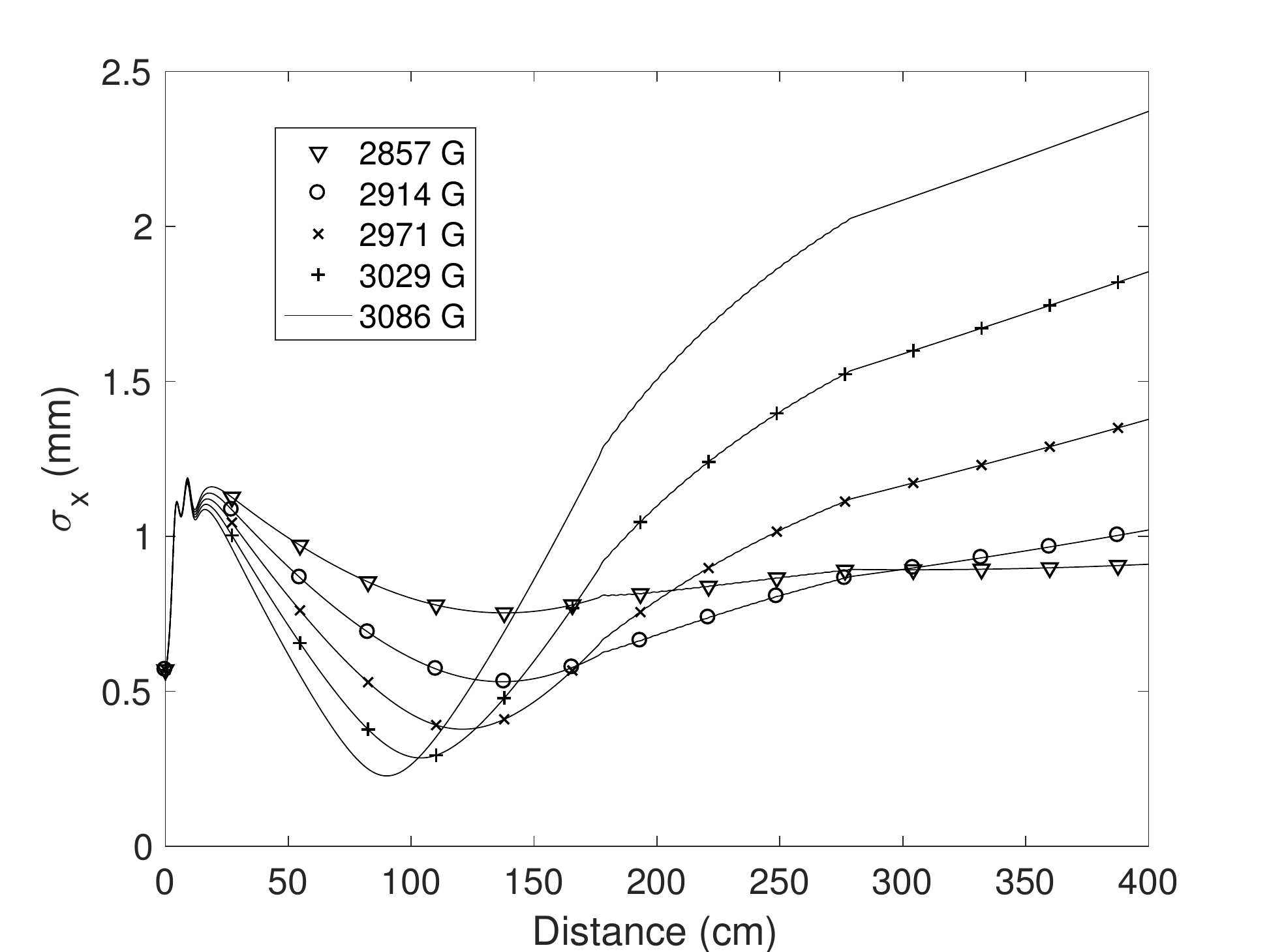}}  \\
\subfloat[]{\includegraphics[width=0.4\textwidth] {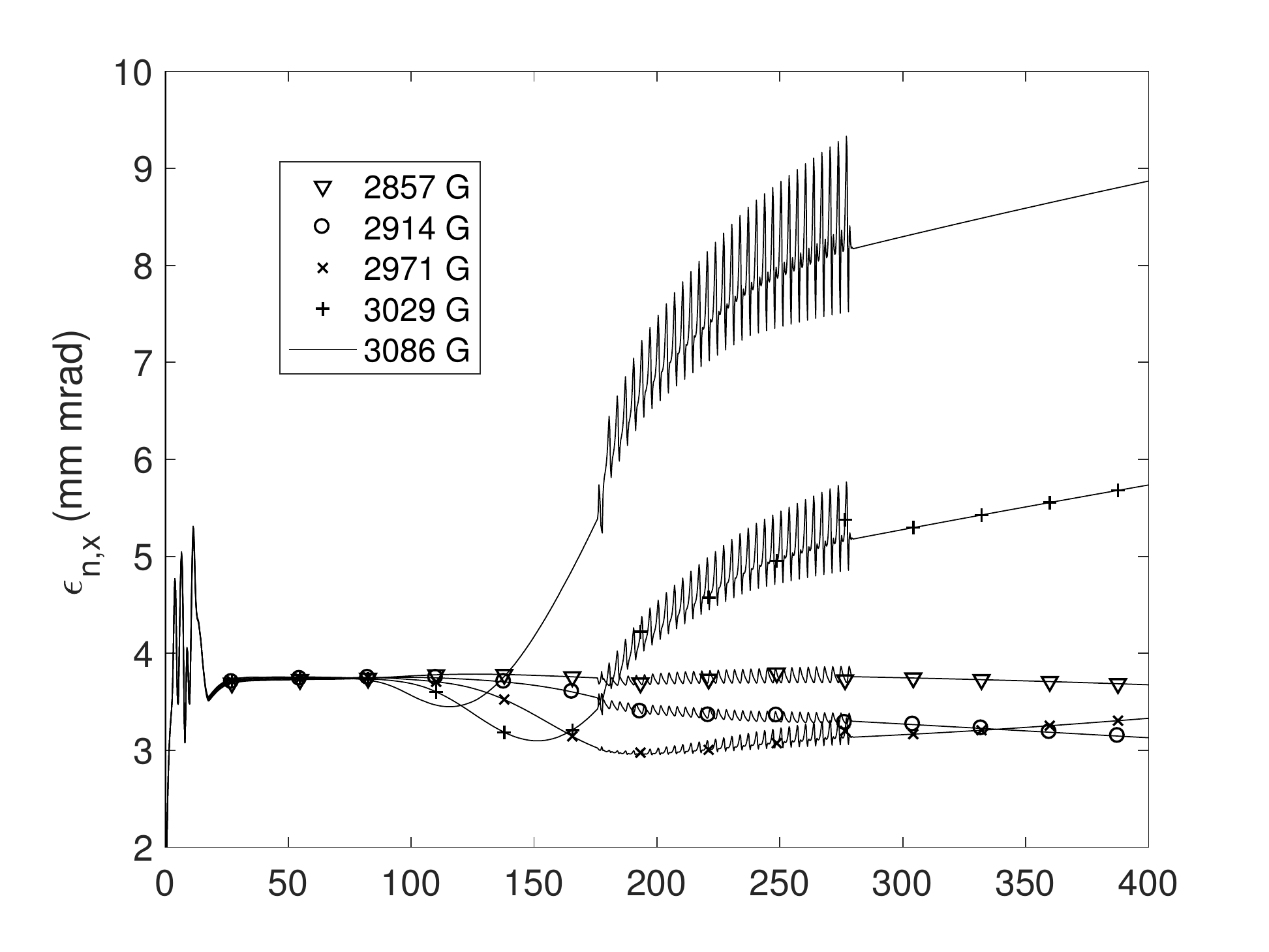}}  
\caption{(a) The ratio of outward pressure due to emittance and the defocusing space charge force, $\varepsilon_n^2\beta\gamma / \sigma^2 \kappa_s$, for a 1$\,$nC beam at the energy of 6.6$\,$MeV under different solenoid fields. (b) Corresponding rms beam size, $\sigma_x$, and (c) emittance, $\varepsilon_{n,x}$, curves. In (a) solid vertical line denotes the location of the pepper-pot mask whereas dashed vertical lines mark the exit of the gun (where also the focusing solenoid is located), entrance and exit of the linac, respectively.}
\label{fig:spacecharge}
\end{figure}

For a relativistic beam ($\beta\approx1$), apart from the constant normalised beam energy $\gamma$, and beam current $I$, the ratio of emittance and space charge defocusing scales proportional to the square of the beam emittance and inversely proportional to the square of the rms beam size. Hence the evolution of this ratio is mainly determined by the beam envelope. Figure \ref{fig:spacecharge}-(a) shows that the ratio, $\varepsilon_n^2\beta\gamma / \sigma^2 \kappa_s$ evolves through the beamline for varying $\varepsilon_n$ and $\sigma$(s) (given in Fig. \ref{fig:spacecharge}-(b)) as well as $\beta$(s) and $\gamma$(s) as beam undergoes focusing and acceleration. Here, `s' is the curvilinear coordinate following the beam trajectory. In Fig. \ref{fig:spacecharge}-(a), the vertical dashed lines comprise the regions of acceleration due to the RF gun and the booster linac. In these regions the emittance oscillations are visible as expected due to the time varying nature of the RF fields and envelope mismatch \cite{Serafini}, hence, the amplitude of these oscillations depends on solenoidal focusing as well as bunch charge (strength of the space charge force, not studies here). The horizontal red dashed line indicates the value where the space charge defocusing and intrinsic beam emittance are equal. Consequently, beam configurations below this point lead to space charge dominated beams. The solid black line marks the location of the emittance measurement (location of the mask at 132$\,$cm). 

Figure \ref{fig:spacecharge}-(a) shows that for all the studied focusing conditions the ratio $\epsilon^2_n \beta \gamma / \sigma^2 \kappa_s$ remains less than one throughout the majority of the beamline (below the red line marking the space charge limit). The evolution of a beam envelope accelerated under such conditions will therefore remain dominated by space charge forces. Consequently, to measure only the intrinsic beam emittance, a method to remove the space charge contribution to the evolving beam envelope is required. A common technique is the pepper-pot method, which uses a mask to isolate the intrinsic beam emittance, and captures the 4D transverse phase-space ($x$, $x^{\prime}$, $y$, $y^{\prime}$) of a beam in a single shot \cite{Zhang, Anderson}. The beam envelopes and projected emittance evolution are presented in \ref{fig:spacecharge}-(b) with corresponding solenoid field values. Emittance compensation occurs at 2914$\,$G minimising the emittance delivered after the linac \cite{Serafini, Floettmann}.

\section{4D Phase Space Sampling}
\label{sampling}
In previous section we show that for sufficiently intense beams at low energy, emittance should be measured under conditions where the defocusing due to space charge is eliminated and hence only defocusing due to beam emittance is detected. This is facilitated with a mask, comprised of either slits or holes in some high-Z material, and is designed to ensure that having passed through, each electron ensemble (beam-let) undergoes negligible space charge defocusing. In the case of a mask with slits, the measurement is only in one plane (x-x$^{\prime}$ or y-y$^{\prime}$ depending on the orientation of the slits) however utilising a mask with holes, which is called a "pepper-pot" mask,  allows four dimensional (x-x$^{\prime}$ and y-y$^{\prime}$ simultaneously) single shot measurement of the transverse emittance.

\subsection{Emittance Analysis Algorithm for Pepper-pot Method}
\label{algorithm}
\begin{figure}[htb!] 
\centering
\includegraphics[width=0.4\textwidth] {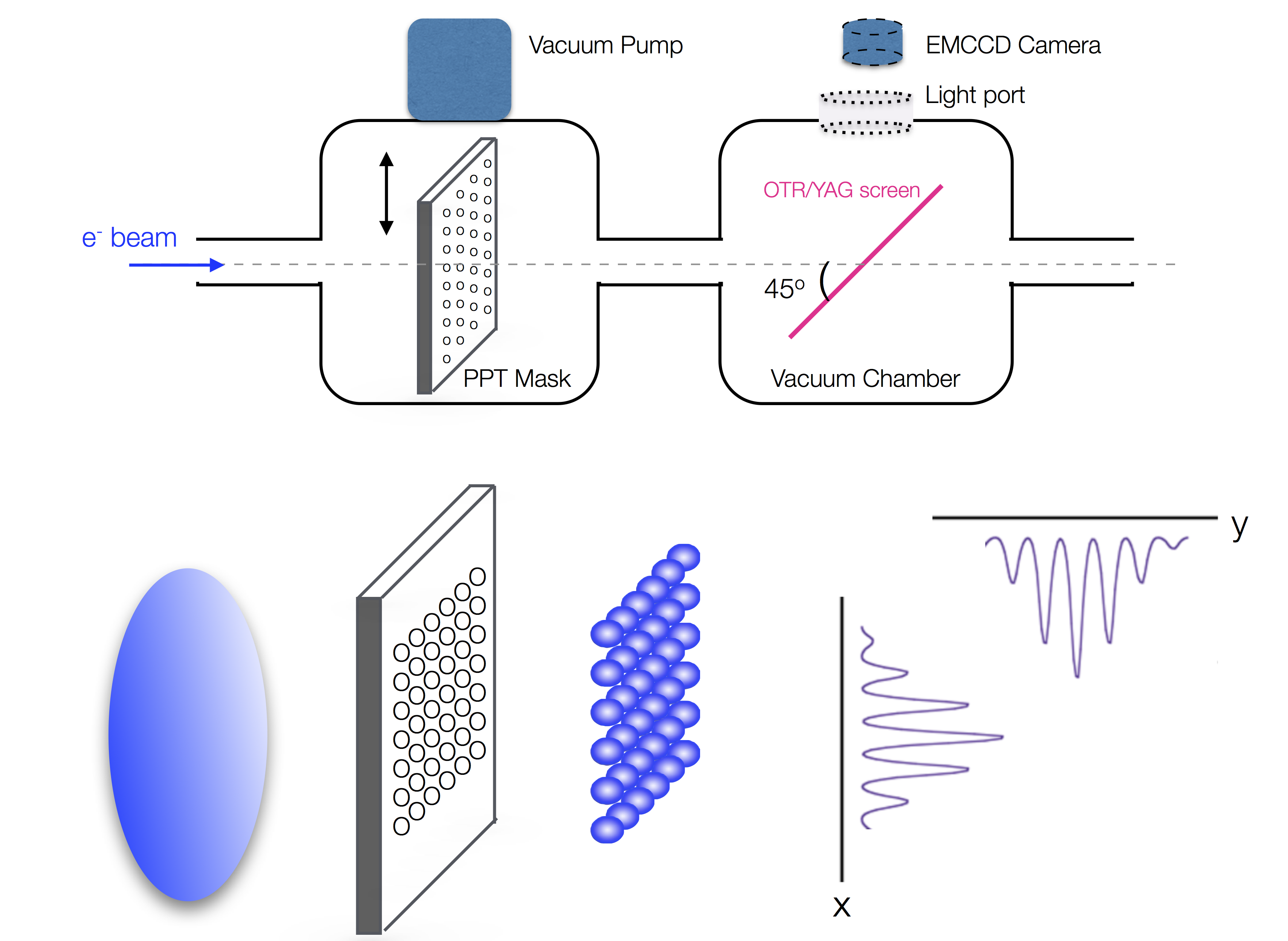} 
\captionsetup{justification=centering}
\caption{The working principle of the pepper-pot emittance measurement.}
\label{fig:ppot}
\end{figure}

This method suggests to split a space charge dominated beam into beam-lets, by using a mask with holes arranged in a rectangular matrix, so that each beam-let carries an amount of charge to pose no significant space charge defocusing \cite{Zhang, Anderson}. After a certain propagation through a drift section, these beam-lets can be observed on a fluorescent or optical transition radiation screen located downstream of the mask as sketched in Fig. \ref{fig:ppot}. Intensity projection of the beam-lets  on either axis are used to calculate each term in the rms emittance equation as a weighted sum over the relative intensities of individual beam-lets as shown from Eq. \ref{eqn:emitt_formulae} to Eq. \ref{eqn:xxprime} \cite{Anderson}. 

\begin{equation}
\varepsilon_x = \sqrt{\langle x^2 \rangle \langle x^{\prime 2} \rangle - \langle xx^{\prime} \rangle ^2} 
\label{eqn:emitt_formulae}
\end{equation}
\begin{equation}
\langle x^2 \rangle = \frac{\sum_{i=1}^N \rho_i x_{i,c}^2}{\sum_{i=1}^N \rho_i }
\label{eqn:xsquare}
\end{equation}
\begin{equation}
\langle x^{\prime 2} \rangle =  \frac{ \sum_{i=1}^N \rho_i (x^{\prime 2}_{i,c} + \sigma_i^{\prime 2}) }{\sum_{i=1}^N \rho_i }
\label{eqn:x2prime}
\end{equation}
\begin{equation}
\langle xx^{\prime} \rangle = \frac{ \sum_{i=1}^N \rho_i x_{i,c} x_{i,c}^{\prime}}{\sum_{i=1}^N \rho_i }
\label{eqn:xxprime}
\end{equation}

In the above equations, $c$ in the indices denotes that the values are with respect to the centroid of the beam-lets (in the analysis code centroid is defined as the beam-let with the largest intensity); $\rho_i$ is the measured  intensity of the $i^{th}$ beam-let;  $\sigma^{\prime}_i = \sigma_i/L$ is the spread in the divergence of the $i^{th}$ beam-let due to the finite beamlet width ($\sigma_i$); and $x^{\prime}_i$ is the divergence of the $i^{th}$ beam-let calculated by correlating the hole locations and beam-let locations on the screen as $\langle x_i - id \rangle / L$, where d is the distance between the holes.

\subsection{Mask Design}
\label{design}
The basic design criteria for a pepper-pot system are formed considering the strength of space charge force present, divergence and rms size of the beam, and the angular resolution of the system \cite{Piot}. After the mask beam-lets should carry a fraction of the initial charge to prevent them from exerting any space charge force. The ratio between the space charge and the beam emittance, $R^{\prime}$, can be written in terms of the mask geometry, i.e., $\omega$ (hole diameter), $d$ (centre-to-centre distance between the holes) and $L$ (mask-screen distance) (Eq. \ref{eqn:rprime}) \cite{Anderson}. f
\begin{equation}
R^{\prime} = \frac{2I}{\gamma^2 I_0 \varepsilon_n} \frac{\omega L}{d}
\label{eqn:rprime}
\end{equation}

The thickness of the mask is chosen so that either all particles outside the hole apertures of the mask are completely absorbed or they are scattered to form a distinctively separate signal to that originated from the beam-lets. To eliminate electrons diffusing the mask, one should chose a mask thickness of at least one radiation length. The radiation length, $L_s$, of the material for a given incoming beam energy, $E$, is a guideline to determine mask thickness where $\rho$ is the density of mask material (19.25$\,g/cm^3$ for Tungsten).  It is calculated using the stopping power regarding the target material, using the Bethe-Bloch equation, $dE/dx$, and is given in practical units as in Eq. \ref{eqn:Ls} \cite{Anderson}. 
\begin{equation}
L_s = \frac{E}{\frac{dE}{dx}} \approx \frac{E(MeV)}{1.5(MeV\,cm^2\,g^{-1})\rho(g\,cm^{-3})}
\label{eqn:Ls}
\end{equation}

When the expected operational intensity is high enough to not raise any concern on collecting enough light for imaging, then the mask can be made one radiation length so that no background due to electrons diffusing is generated. On the other hand, the aspect ratio of the mask (ratio of hole sizes to mask thickness) can be challenging to achieve for very fine matrix of holes for thick masks. The most common machining technique is the laser drilling which is limited about 200$\,\mu$m for the achievable distance between holes and aspect ratio of the mask, i.e., the ratio between the mask thickness and the hole diameter (generally up to 110$\%$ is possible). Electrical discharge machining is another existing technique with finer machining capabilities for a larger cost and can easily overcome the machining limitations mentioned above. When there are machining limitations, a pepper-pot mask thinner than the radiation length can provide more flexibility. In this case, electrons masked by Tungsten slab will be scattered rather than absorbed. Such scattered electrons produce a generally Gaussian distributed background on top of electrons propagating through the holes which can be differentiated from this beam-let distribution with offline background subtraction. 

Once the beam-lets propagate through the mask, they travel through a certain drift length before reaching the screen. Depending on the beam divergence, $\sigma^{\prime}$, the observation screen should be distanced from the mask to prevent beam-lets from overlapping on the screen, namely, fulfilling the condition $4\sigma^{\prime}L < d$.

Finally, one can ensure that position and angle resolutions are comparable, $\sigma/d = L\sigma^{\prime}/r_d$, where $\sigma$ is the rms beam size and $r_d$ is the resolution of the detector (pixel size of the CCD camera) which is taken 10$\,\mu$m for this study. 

An analytical initial mask design was performed under the focusing condition ensuring that the beam envelope is matched to the linac entrance. A range for mask geometries spanning from 100-200$\,\mu$m for the hole diameter, 100-1250$\,\mu$m for the centre-to-centre hole distance and 30-270$\,$mm for the mask-to-screen distance were explored in the light of the design criteria above.  The results fulfilling $R^{\prime} \approx 1$ are summarised in Table \ref{tbl:ana_mask_design}. Apart from the criteria summarised in the table, one should make sure that the angular aperture of a single hole on the mask, $\omega/4L_s$ is larger than the expected beam angle, $\varepsilon_n/\gamma \sigma$. For the parameters considered in this study, the average beam angle ranges from 0.5 to $<$1$\,$mrad while the angular aperture of the mask ranges from 10 to 22$\,$mrad depending on the hole diameter and assuming that the mask thickness is equal to the radiation length (2.3$\,$mm) of Tungsten for 6.6$\,$MeV electrons. Therefore the angular beam clearance is more than sufficient for the geometries considered.
\begin{table}[hbt!] 
\centering
\small
\caption{A summary of semi-analytical mask designs regarding the criteria explained in Section \ref{design}.} 
\resizebox{8cm}{!}
	{ 
   		\begin{tabular}{lcccccc}
       		\hline
 $\omega$ ($\mu$m) \hspace{0.5em} & d ($\mu$m) \hspace{0.5em} & L (mm) \hspace{0.5em} & R$^{\prime}$ \hspace{0.5em} & 4$\sigma^{\prime}L$ ($\mu$m) \hspace{0.5em}  &  $\sigma$/d  \hspace{0.5em} &  L$\sigma^{\prime}/r_d$  \\   
		\hline
       		\hline
		100	&	100	&	30	&	0.9	&	60	&	5.3	&	1.5\\
		100	&	200	&	60	&	0.9	&	120	&	2.6	&	3.0\\
		100	&	300	&	95	&	1.0	&	190	&	1.8	&	4.7\\
		100	&	400	&	120	&	0.9	&	240	&	1.3	&	6.0\\
		100	&	450	&	140	&	1.0	&	280	&	1.2	&	7.0\\
		100	&	500	&	155	&	1.0	&	310	&	1.1	&	7.7\\
		100	&	600	&	190	&	1.0	&	380	&	0.9	&	9.5\\
		200	&	850	&	135	&	1.0	&	270	&	0.6	&	6.7\\
		200	&	1250	&	195	&	1.0	&	390	&	0.4	&	9.7\\
		100	&	650	&	200	&	1.0	&	400	&	0.8	&	10.0\\
		100	&	850	&	270	&	1.0	&	540	&	0.6	&	13.5\\
		150	&	650	&	135	&	1.0	&	270	&	0.8	&	6.7\\
		150	&	1250	&	265	&	1.0	&	530	&	0.4	&	13.2\\
		150	&	950	&	200	&	1.0	&	400	&	0.6	&	10.0\\

       		\hline		
   		\end{tabular}
   	}
\label{tbl:ana_mask_design}
\end{table}

In the table, the values for $4\sigma^{\prime}L$ suggests no beam-let overlap on the screen. However, the position resolution $\sigma/d$ values become an order of magnitude smaller than the angular resolution, $L\sigma^{\prime}/r_d$, towards the bottom of the table. This might imply one might not achieve enough number of beam-lets on the screen. These design points were tested for the beam distributions created by PARMELA for the focusing conditions given in Section \ref{characteristics} using a custom tracking code introduced in the next section.  

\subsection{Tracking Simulations}
\label{tracking}
An initial Gaussian beam with 0.5$\,$mm radius and 4$\,$ps bunch length was tracked with PARMELA in the presence of previously mentioned RF and magnetic fields. The distributions at the location of the mask are retrieved from PARMELA and further tracked through a mask with a given geometry up to a screen located downstream with a MATLAB \cite{MATLAB} script. Once the distributions on the screen are extracted, after a polynomial background subtraction, each beam-let is processed to calculate the normalised emittance of the initial beam entering the mask using pepper-pot algorithm introduced in Section \ref{algorithm}. One should note that, in a real life measurement the background can be more complicated due to external effects  such as any ambient light, heating of the screen and x-rays due to the interaction of the electrons with Tungsten mask. 

PARMELA simulates macro particles which are ensembles representing many real particles. A macro particle unfolding algorithm is included to provide a more realistic signal intensity for tracking. This algorithm creates a number of new particles in a Gaussian distribution within a certain radius around the mean position of each macro particle with the same divergence of this mother particle. In this study, each macro-particle is unfolded into 100 new particles within a radius of 10$\,\mu$m. The unfolding method smooths out the distributions to help with the tracking and analysis. Subsequent tests showed that the final normalised emittance result changes 1-2$\%$ between unfolding with 100 new particles and no unfolding.
 
Unfolded particles coinciding with the holes on the front face of the mask are considered as survived particles and kept for further analysis. Divergence of the beam is taken into account during the propagation through the mask thickness and the drift section up to the screen; a particle with a trajectory exceeding a hole aperture is considered as an absorbed particle and is removed from the rest of the calculations. 

\begin{figure*}[htb!] 
\centering
\subfloat[]{\includegraphics[width=0.45\textwidth] {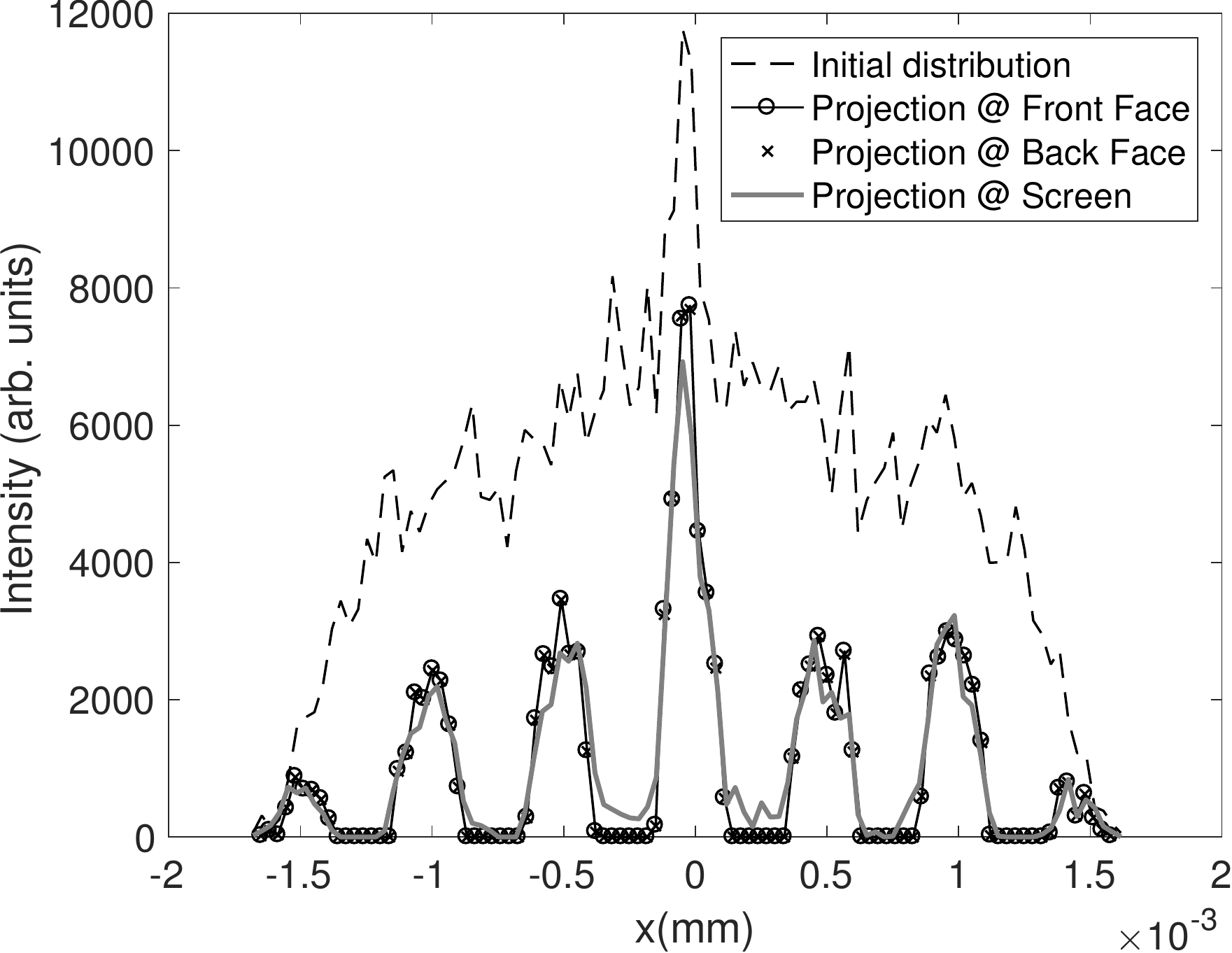}} \hspace{0.5em}
\subfloat[]{\includegraphics[width=0.44\textwidth] {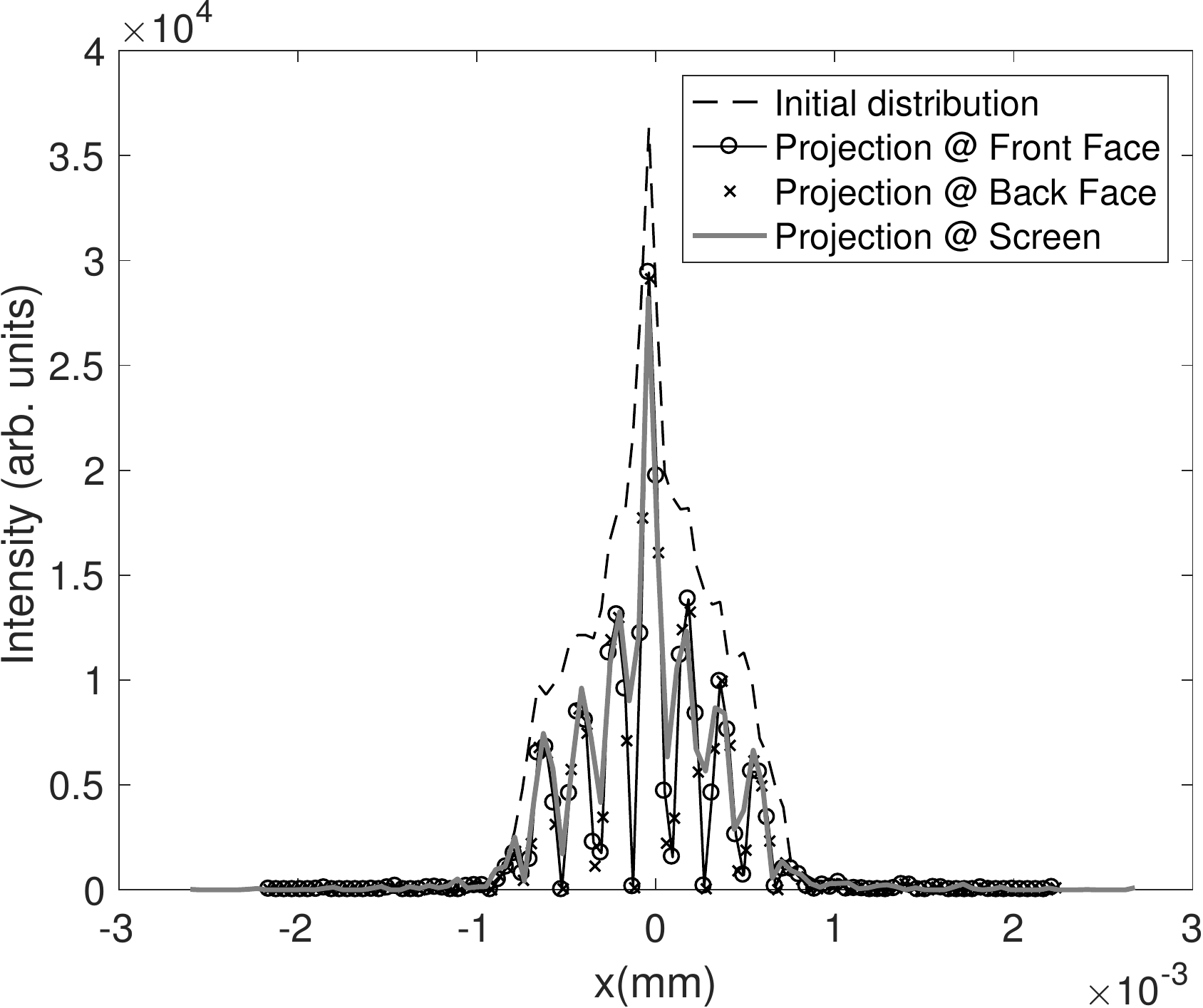}} \\
\subfloat[]{\includegraphics[width=0.43\textwidth] {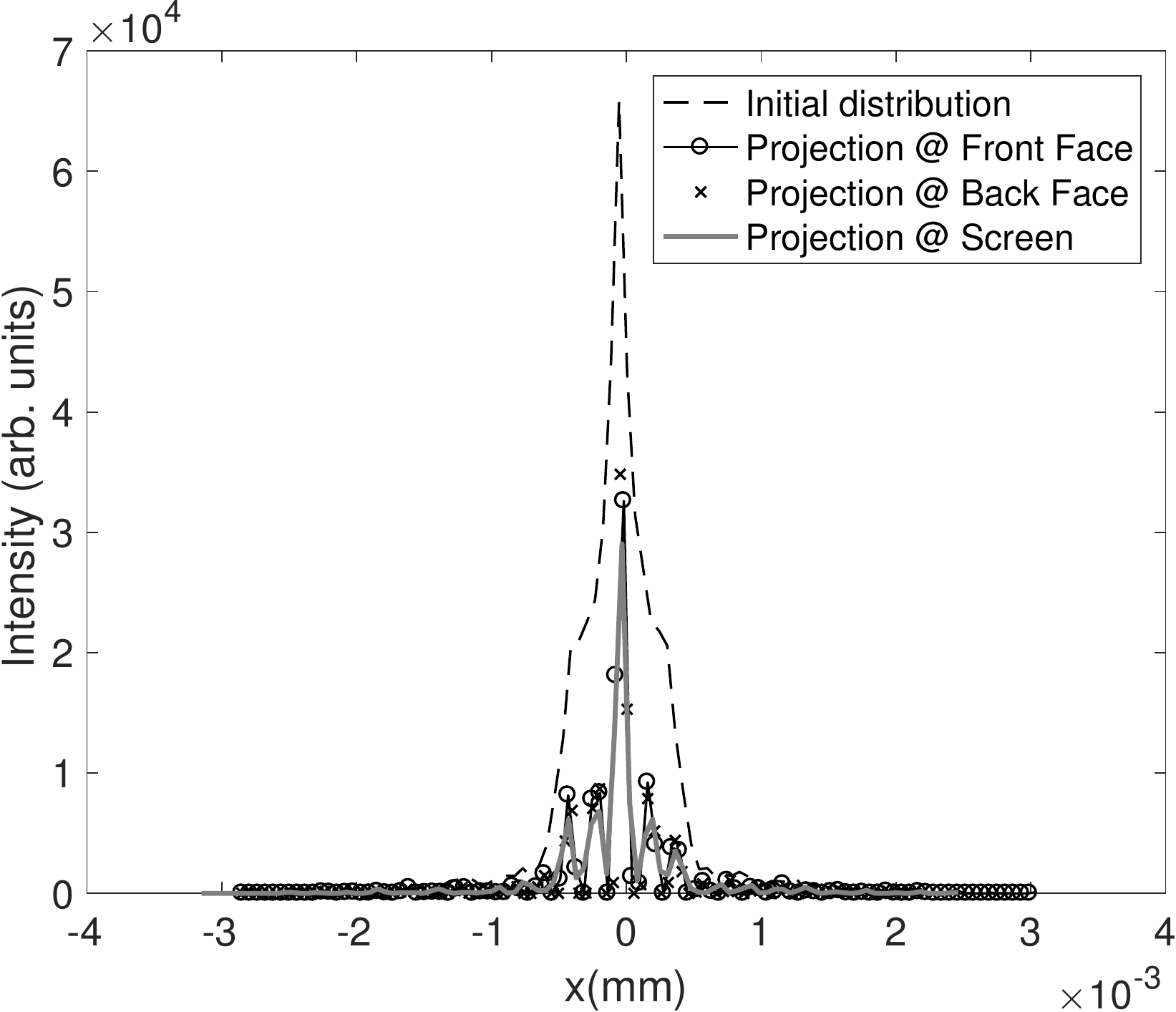}}  \hspace{0.5em}
\subfloat[]{\includegraphics[width=0.43\textwidth] {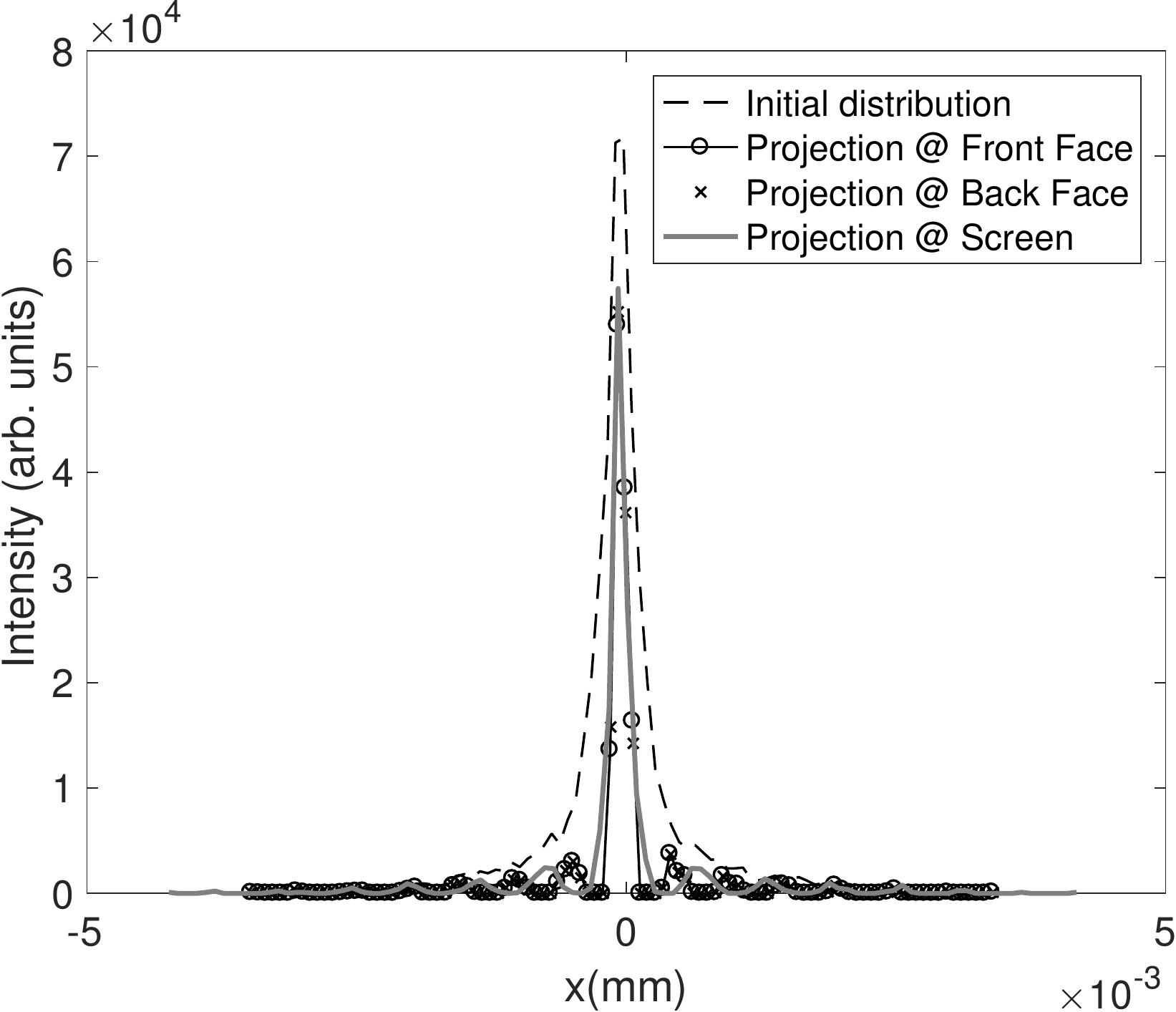}} 
\caption{Transverse projections of the initial particle distributions (dashed lines) at the location of the mask (132$\,$cm away from the cathode located at 0$\,$cm), the front (depicted by circles) and back faces (depicted by crosses) of the mask and finally at the screen (thick grey line), for a focusing field provided by a solenoid current of (a) 2857$\,$G (under-focused), (b) 2914$\,$G (envelope matched to linac), (c) 3029$\,$G (minimum emittance at the mask) and (d) 3086$\,$G (over-focused). Initial distribution in the legends refers to the incident distribution at the mask location.}
\label{fig:results}
\end{figure*}

Results of beam tracking for four different focusing conditions are presented in Fig. \ref{fig:results}. These include the transverse projections of initial distributions incoming at the mask location, distributions at the front and back face of the mask and the distributions at the screen. From Fig. \ref{fig:results}-(a) to (d), distributions represent an under-focused envelope, an envelope satisfying the matching with linac, an envelope minimising the emittance at the mask and an over-focused envelope. 

The transmission to the screen is found to be 10 to 50$\%$ depending on the number and size, namely the density, of the holes on the mask. The tracking model assumes a mask thickness of more than or equal to the radiation length of the material. Hence particles hitting the mask are considered as absorbed by the mask rather than propagating through, scattering from the mask material and hitting the screen. Under this assumption, 1$\%$ of the particles travelling through the holes are absorbed. In reference \cite{Cheymol} the scattering of particles at the edges of a slit is studied via Monte Carlo simulations. These results are used to investigate the effects of slit geometry as well as the mask material on the amount of scattered background particles reaching the observation screen in investigated. Results reported in this paper are based on a mask design with straight cylindrical holes. 

\subsection{Systematic Error Analysis}
The position ($\sigma_{x_i}$) and intensity ($\sigma_{\rho_i}$) stability of the beam-lets can be approximated to the pointing and intensity stability of the laser, respectively, by ignoring any field errors or misalignment. The error on the angles is calculated using the beam-let widths and distance between the mask and the observation screen as in Eq. \ref{eqn:x2prime}, $\sigma^{\prime} = \sigma_i / L$. 

Once this average systematic error is much smaller than the statistical error on the measurements, the diagnostics is deemed accurate within the given beam stability.

The systematic error on a function $f(x,y,z)$ is derived using the general expression given in Eq. \ref{eqn:sys_err0},
\begin{equation}
\sigma^2_{f(x,y,z)} = \sum\bigg(\frac{\partial f(x,y,z)}{\partial i}\bigg)^2\sigma_i^2
\label{eqn:sys_err0}
\end{equation}
where $\sigma_i=\sigma_x, \sigma_y, \sigma_z$ are the uncertainties on observables with $i=x,y,z$. 

The systematic errors such as the position and angle resolution of the mask ($\sigma_x$,$\sigma_{x^{\prime}}$), as well as the intensity fluctuations of the beam ($\sigma_{\rho}$), were propagated through the variables of each term in the rms emittance calculation in previous section and are presented through Eq. \ref{eqn:sys_err1}-\ref{eqn:sys_err2}. 

\begin{equation}
\sigma^2_{\sum\rho x^2} = \sum^N_{i=1}x^4_i\sigma^2_{\rho_i} + 4x^2_i\rho^2_i\sigma^2_{x_i}
\label{eqn:sys_err1}
\end{equation} 

\begin{equation}
\sigma^2_{\sum \rho x^{\prime2}} = \sum^N_{i=1}x_i^{\prime4} \sigma^2_{\rho_i}+4x^2_i\rho^2_i\sigma^2_{x_i}
\end{equation} 

\begin{equation}
\sigma^2_{\sum\rho x x^{\prime}} = \sum^N_{i=1}\rho^2_ix_i^{\prime}\sigma^2_{x_i} + \rho^2_ix^2_i\sigma^2_{x_i^{\prime}} + x_i^2 x_i^{\prime 2}\sigma_{\rho_i}^2
\end{equation} 

\begin{equation}
\sigma^2_{(\sum\rho)^2} = 4\bigg( \sum_{i=1}^N \sigma_{\rho_i}^2 \bigg) \bigg( \sum_{i=1}^N \rho_i \bigg)^2
\label{eqn:sys_err2}
\end{equation} 

Finally, errors associated with each term were combined to provide the systematic error on the emittance measurement. This is shown in Eq. \ref{eqn:sys_error} in terms of the moments and intensities of the beam-lets as well as the individual errors on these observables.

\begin{equation*}
 \sigma^2_{\varepsilon} =  \sigma^2_{(\sum\rho x^2 \sum\rho x^{\prime 2}-(\sum\rho x x^{\prime})^2)/\sum \rho} 
\end{equation*}

\begin{equation*}
  = \frac{(\rho_ix_i^{\prime 2})^2(4\rho_i^2x_i^2\sigma^2_{x_i}+x_i^4\sigma^2_{\rho_i})}{4\varepsilon^2(\sum^N_{i=1}\rho_i)^4} 
 \end{equation*}

 \begin{equation*}
+\frac{(\rho_i^2 x_i^2)^2 (4\rho_i^2 x_i^{\prime 2} \sigma^2_{x_i^{\prime}} + x_i^{\prime 4} \sigma^2_{\rho_i})}{4\varepsilon^2(\sum^N_{i=1}\rho_i)^4} 			     
\end{equation*}

 \begin{equation*}
 +\frac{(\rho_i^2 x_i^{\prime 2}\sigma^2_{x_i} + \rho_i^2 x_i^2 \sigma^2_{x_i^{\prime}} + x_i^2 x_i^{\prime 2} \sigma^2_{\rho_i})(\rho_i x_i x_i^{\prime})^2}{\varepsilon^2(\sum^N_{i=1}\rho_i)^4} 	
 \end{equation*}

 \begin{equation*}
 -\frac{2(\sum^N_{i=1}\rho_i x_i^2)(\sum^N_{i=1}\rho_i x_i^{\prime 2})(\sum^N_{i=1}\rho_i x_i x_i^{\prime})^2(\sum^N_{i=1} \sigma^2_{\rho_i})}{\varepsilon^2(\sum^N_{i=1}\rho_i)^6} 
 \end{equation*}

 \begin{equation}
 +\frac{\varepsilon^2 (\sum^N_{i=1} \sigma^2_{\rho_i})}{(\sum^N_{i=1}\rho_i)^6} 
 \label{eqn:sys_error}
 \end{equation}

One should also note that a correction can be introduced on the angular spread of the beam-lets for beams incident on the mask with large divergences. This is generally a case for electron beams generated through laser-plasma interaction and causes a distortion in beam-let profile during the propagation between the mask and screen which introducing a correlated divergence. The correction used to remove this correlated divergence is given in Eq. \ref{eqn:correction} and the concept is explained and sketched in detail in \cite{ppt_correction}. Note that the notation in the reference is different to the one in this paper regarding the rms angular spread due to emittance ($\sigma^{\prime}_i$, where i refers to the i$^{th}$ beam-let), the measured rms beam-let width ($\sigma_i$), hole width ($\omega$) and distance between the mask and the screen ($L$).

\begin{equation}
\sigma_i^{\prime2} = \frac{\sigma_i^2 - (M\omega/\sqrt{12})^2}{L^2}
\label{eqn:correction}
\end{equation} 
where M is the magnification ratio defined as $M=(L_g+L)/L_g$. Here $L_g$ is the distance from the electron source where electrons originate to the mask. The factor $1/\sqrt{12}$ is introduced to provide the rms value of the flattop distribution created by the holes.

\begin{figure*}[htb!] 
\subfloat[]{\includegraphics[width=0.3\textwidth] {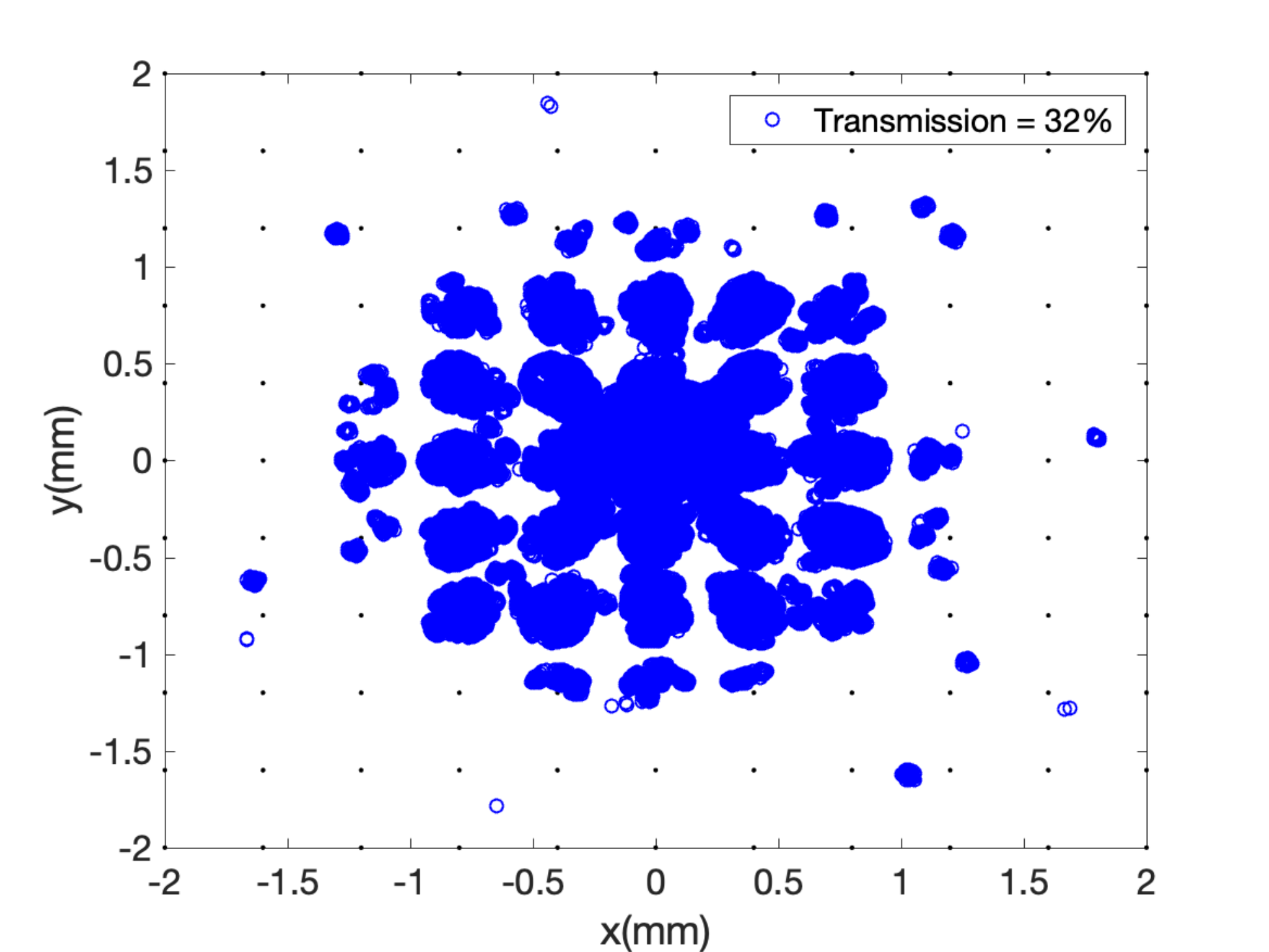}} 
\subfloat[]{\includegraphics[width=0.3\textwidth] {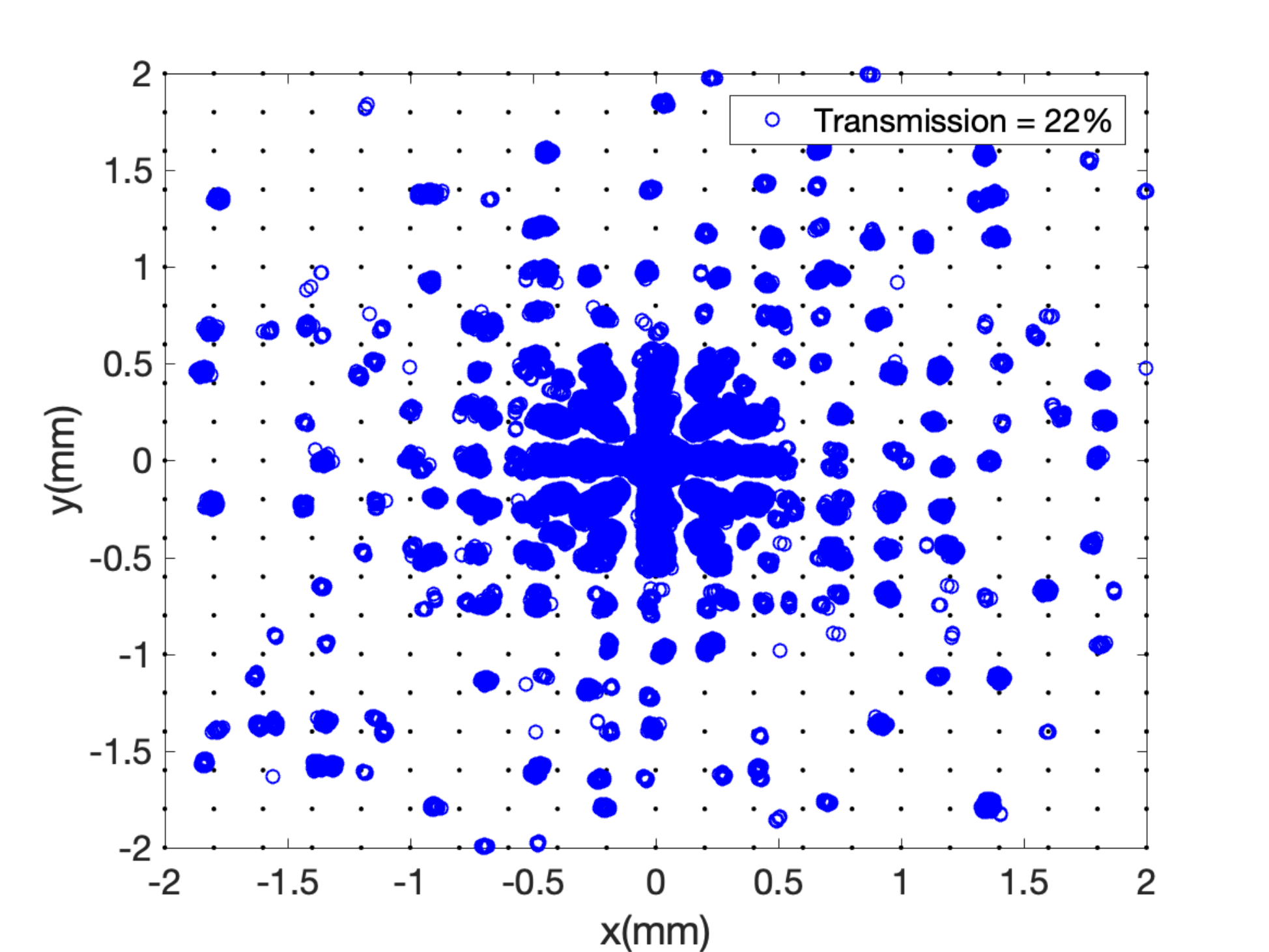}} 
\subfloat[]{\includegraphics[width=0.3\textwidth] {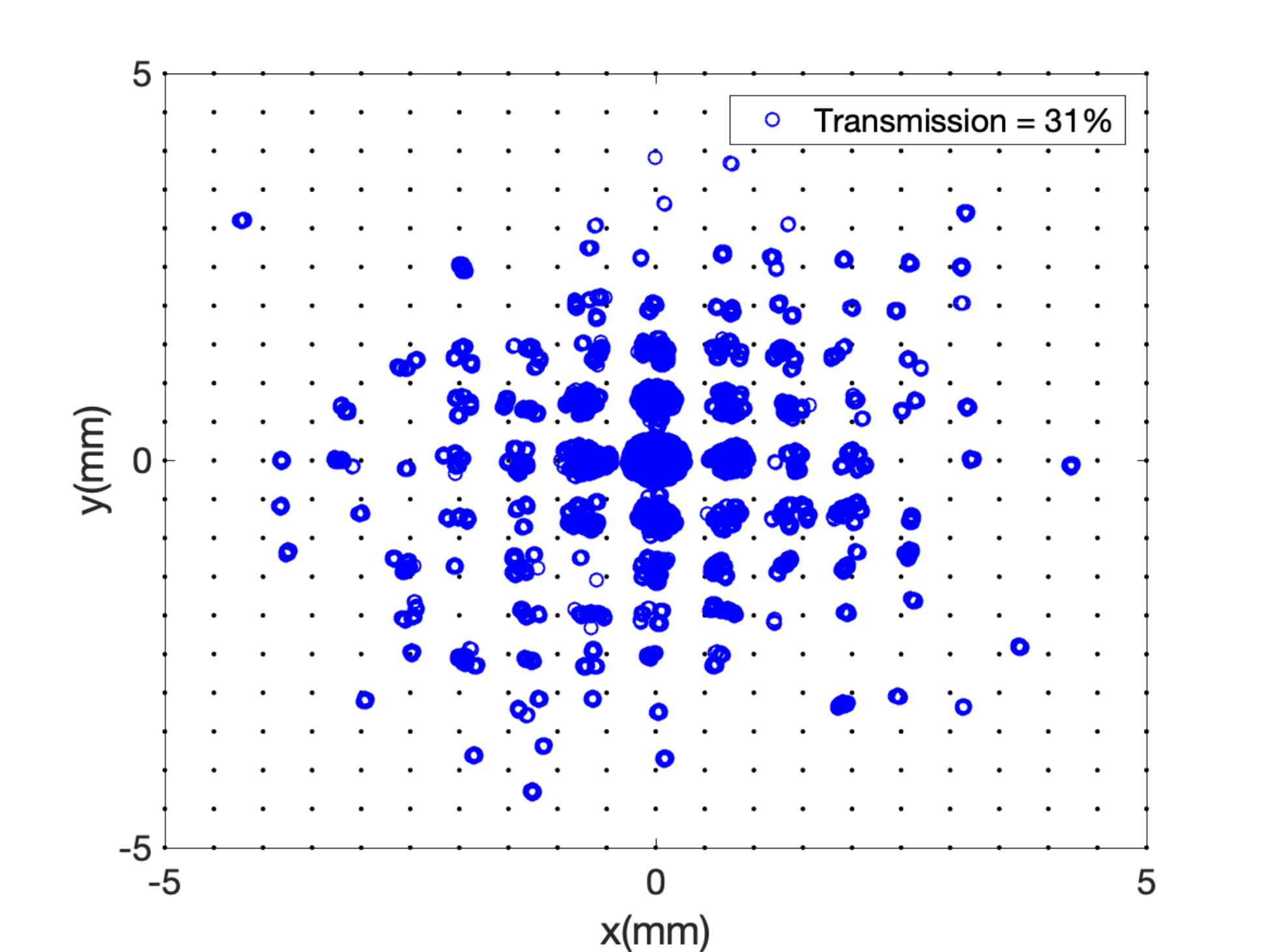}} \\ \vspace{-1.2em}
\subfloat[]{\includegraphics[width=0.3\textwidth] {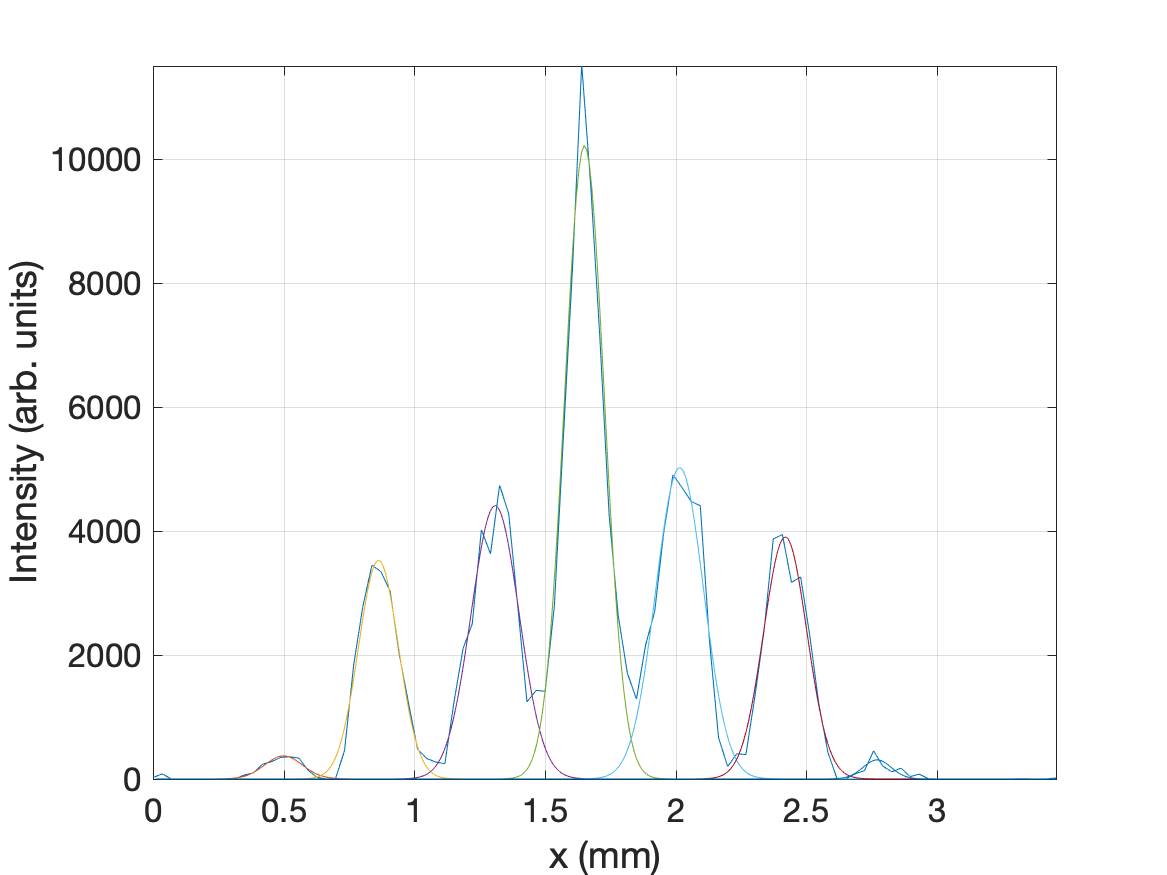}} 
\subfloat[]{\includegraphics[width=0.3\textwidth] {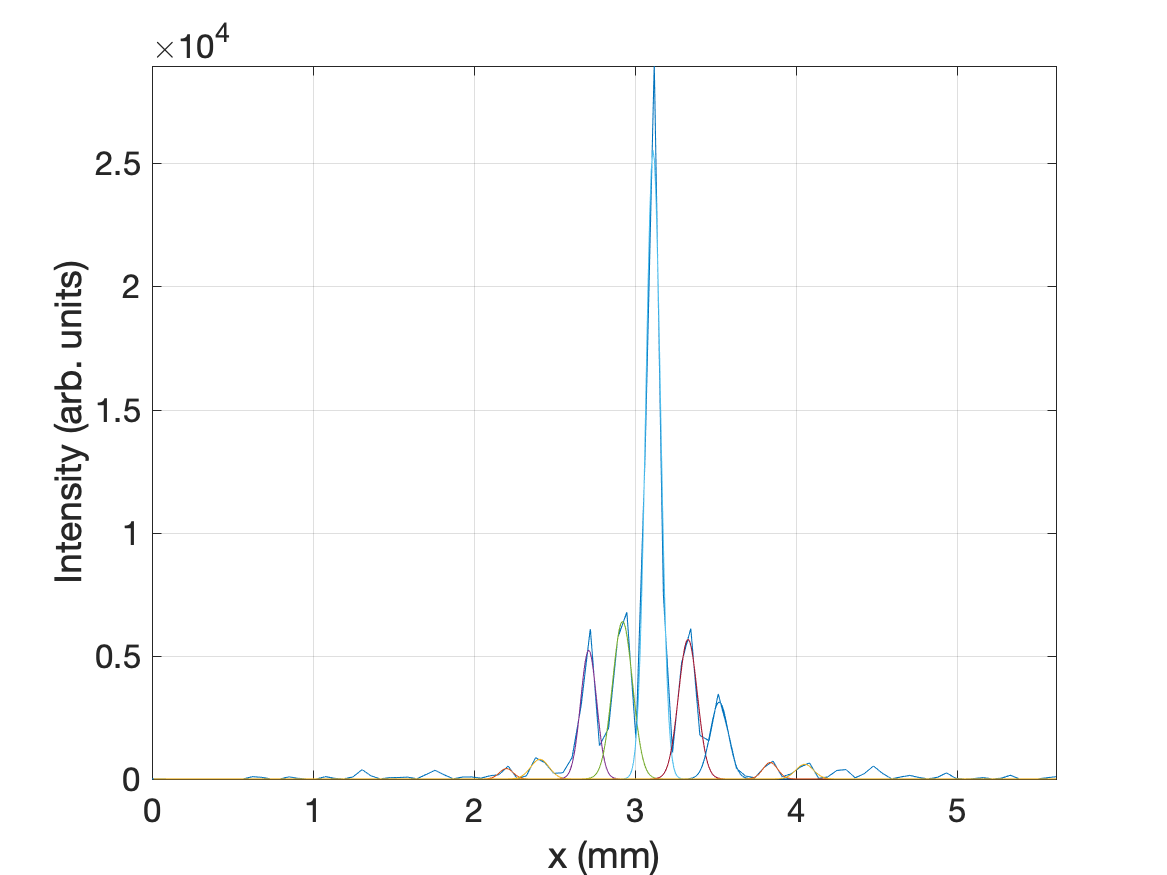}} 
\subfloat[]{\includegraphics[width=0.3\textwidth] {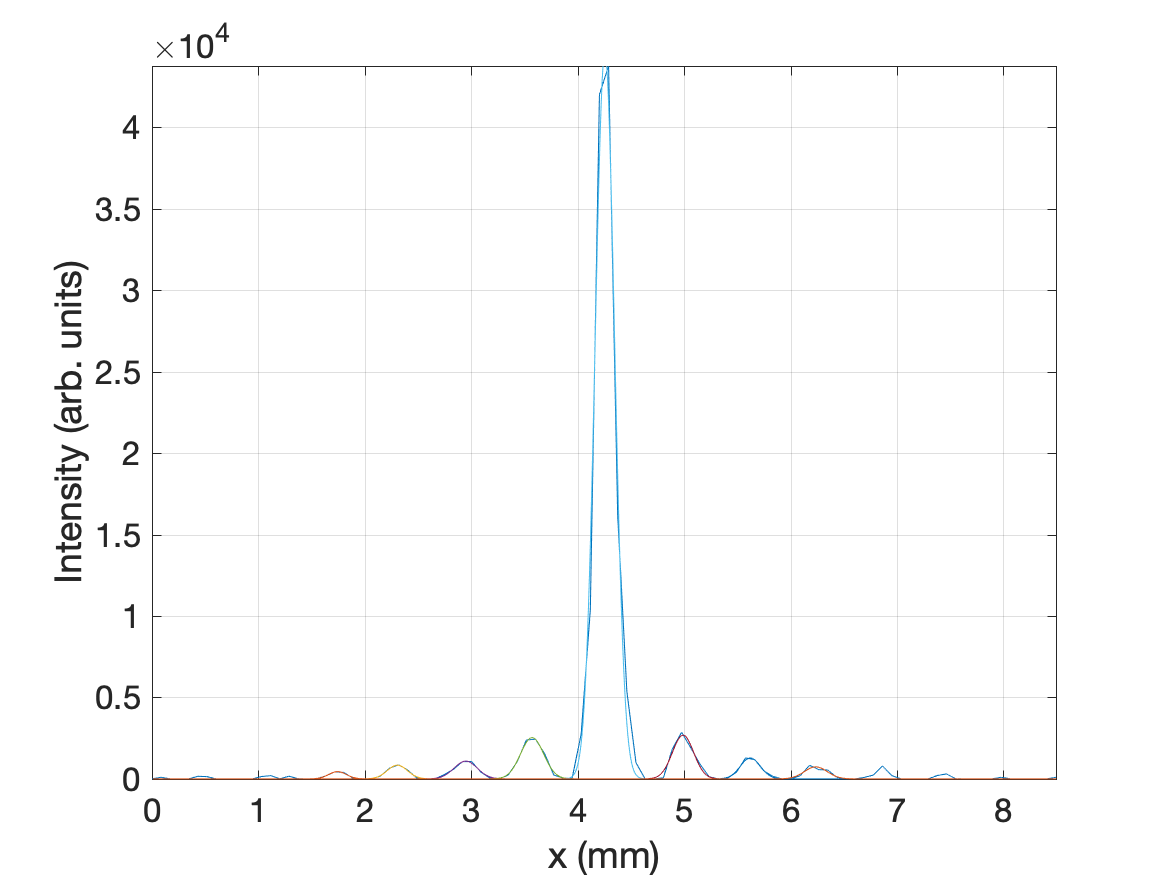}} \\  \vspace{-1.2em}
\subfloat[]{\includegraphics[width=0.3\textwidth] {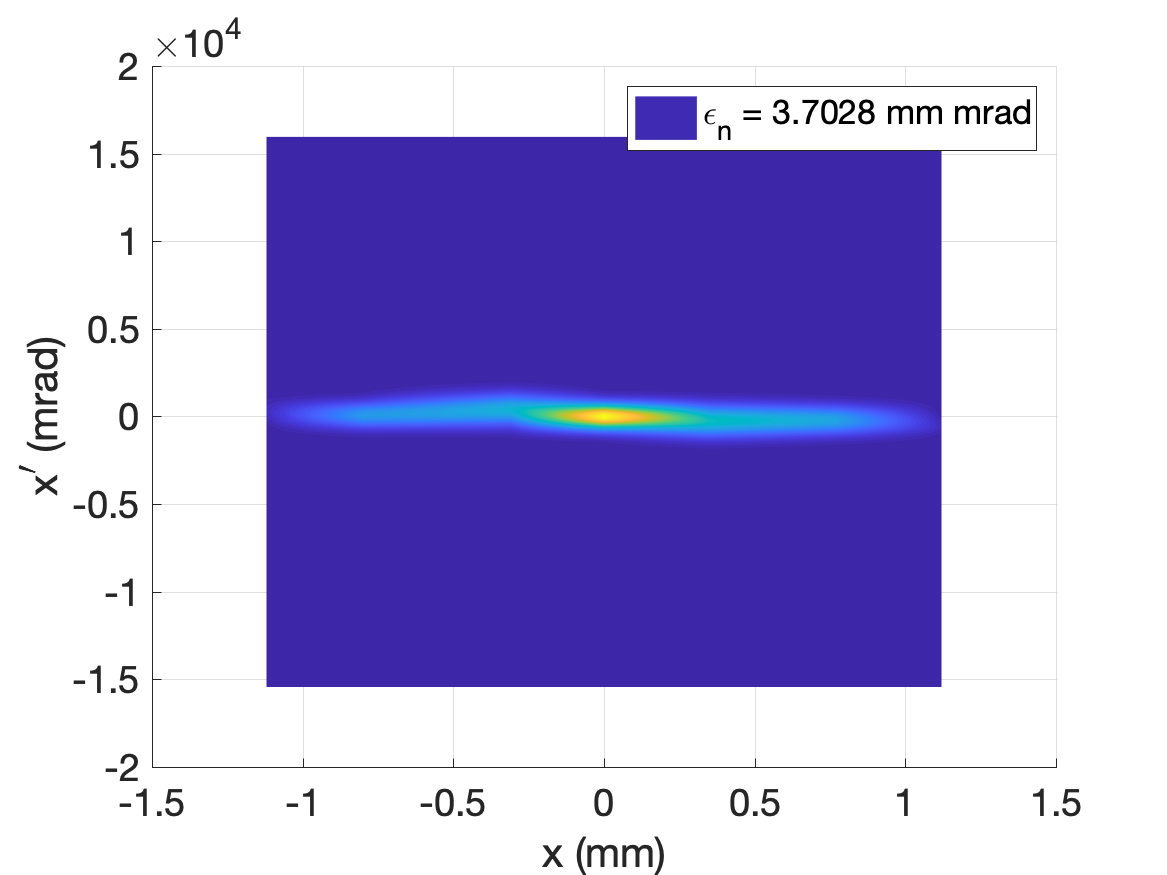}} 
\subfloat[]{\includegraphics[width=0.3\textwidth] {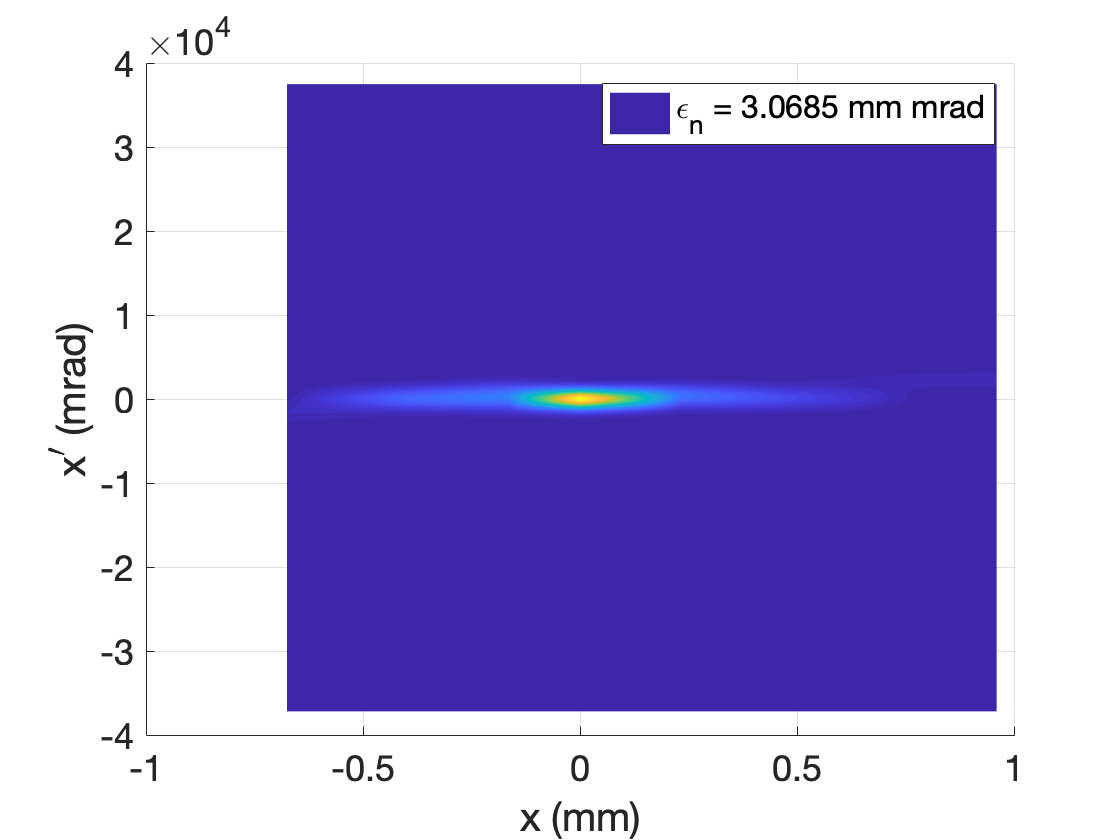}} 
\subfloat[]{\includegraphics[width=0.3\textwidth] {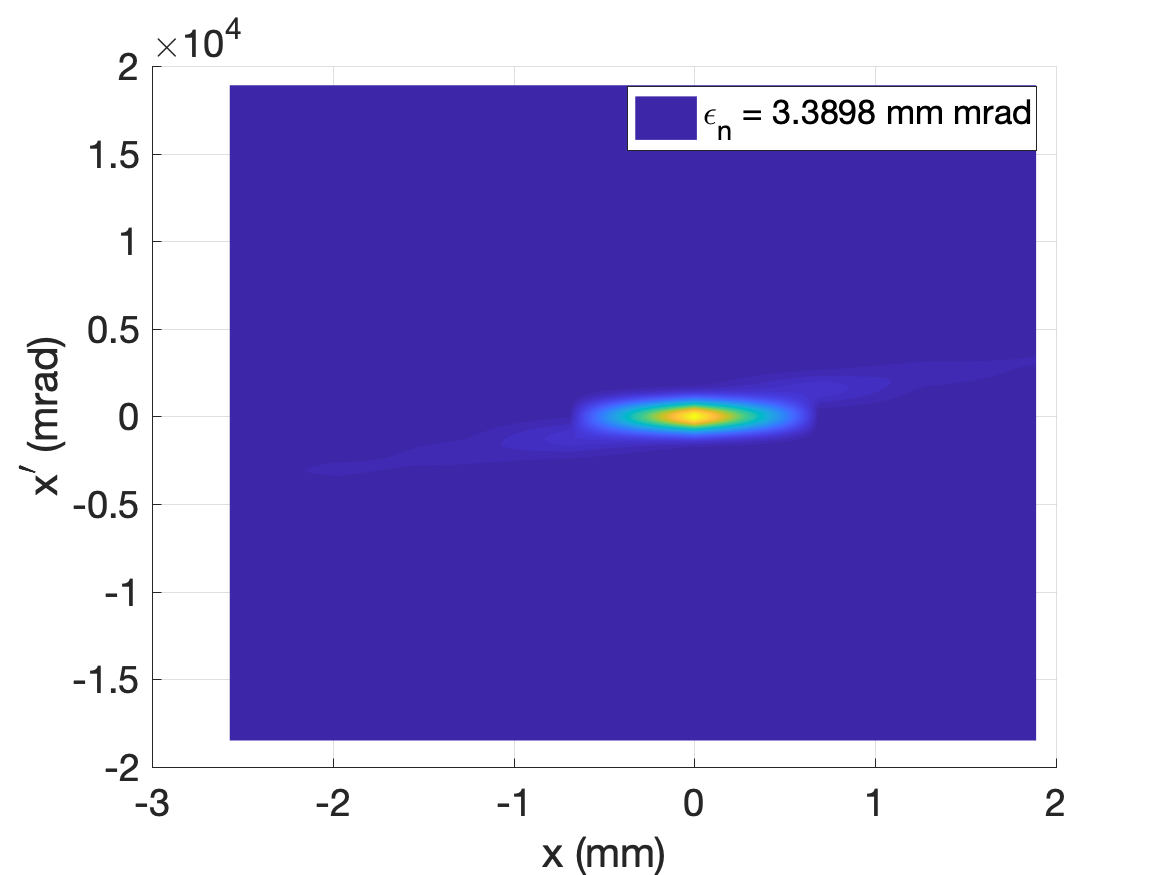}} 
\caption{(Colour online) The simulated beam distributions on a screen (a) 150$\,$mm downstream the mask when the beam envelope is matched to linac, (b) 60$\,$mm downstream the mask for emittance is minimised at this location and (c) 150$\,$mm downstream the mask for an over-focused envelope. Subfigures (d), (e) and (f) present the projections of these particle distributions with Gaussian fit curves on each beam-let. (g), (h), (i) Reconstructed phase space distributions, respectively. The location of the mask is fixed and 132$\,$cm away from the cathode.}
\label{fig:tracking}
\end{figure*}

\section{Performance and Limitations of Emittance Measurement System}
The mask geometries suggested by the analytical calculations were tested using the tracking algorithm through a rigorous iterative process as a function of hole diameter and centre-to-centre hole distances on the mask as well as the distance between the mask and the screen. The mask geometries producing closest results to those retrieved from the PARMELA simulations are presented in Table \ref{tbl:mask_design} in bold characters.

In the weakly focused region (under-focused envelope), the reconstructed emittance values are overestimated  by 40$\%$ in average. A typical example of an under-focused distribution can be seen in Fig. \ref{fig:results}-(a). When under-focused, the projected emittance under space charge force is not compensated (by aligning the phase space angles due to slices). This leads to emittance growth and halo formation, creating high intensity tails at the beam projection at the screen \cite{Wangler}. Furthermore, in the algorithm, given through Eq. \ref{eqn:xsquare} to Eq. \ref{eqn:xxprime}, rms values are calculated as a weighted sum with regards to the relative intensities of the beam-lets. This is compatible with a general case where the beam is Gaussian and outer particles contribute to the rms emittance less than the particles occupying the core of the beam.  Therefore, for non-Gaussian beams the tail particles contribute to the rms emittance as much as the core particles causing an overestimation of the reconstructed emittance through these method. Although the over-focused distributions considered for this study preserve their Gaussian form, by the same token, they are prone to the same problem depending on the level of over-focusing.

This further investigation shows that pepper-pot method is mostly reliant on the incident beam distribution and it works best for beams focused on or at the vicinity of the mask, i.e., beams incident on the mask with a quasi-laminar envelope. The indications of a similar behaviour was observed during the commissioning of PHIN photo-injector which utilised a multi-slit set-up \cite{CLIC}, however the behaviour was not clearly noticeable as measurements were not sufficiently extended outside the matched region.

\begin{table}[hbt!] 
\centering
\small
\caption{The mask geometries originating from the numerical fine-tuning of the initial mask designs suggested by the analytical approach. Emittance reconstruction results for five different focusing conditions introduced in Table \ref{tbl:characteristics}. The bottom line presents the rms normalised emittance values retrieved from PARMELA simulations. The closest values to retrieved ones were marked in bold characters for each case.} 
\resizebox{8cm}{!}
	{ 
   		\begin{tabular}{lcccccccc}
       		\hline
			&		&		&	A	&	B	&	C		&	D	&		E \\
		\hline
		 $\omega$ ($\mu$m) \hspace{0.5em} & d ($\mu$m) \hspace{0.5em} & L (mm) \hspace{0.5em} &  2857$\,$G & 2914$\,$G & 2971$\,$G & 3029$\,$G & 3086$\,$G  \\   
		\hline
       		\hline
		250	&	500	&	150	&	4.95	&	3.82	&		2.75	&	2.55	&		\textbf{3.39}	\\
		200	&	300	&	150	&	4.30	&	3.26	&		2.88	&	3.64	&		4.60	\\
		250	&	400	&	150	&	5.27	&	\textbf{3.70}&	2.98	&	3.72	&		4.01	\\
		250	&	300	&	150	&	5.53	&	4.25	&		3.30	&	4.10	&		5.50	\\
		100	&	200	&	60	&	5.86	&	4.32	&		3.35	&	\textbf{3.10}&	5.54	\\
		150	&	200	&	100	&	5.18	&	3.98	&    \textbf{3.46}&	2.26	&		3.95	\\

       		\hline
		PARMELA & && 3.70&	3.63	&	3.51	&	3.20	&	3.53 \\	
		\hline
   		\end{tabular}
   	}
\label{tbl:mask_design}
\end{table}
\begin{table*}[hbt!] 
\centering
\small
\caption{The results for re-application of analytical criteria explained in Section \ref{design} on mask geometries fined-tuned with numerical optimisation.} 
\resizebox{12cm}{!}
	{ 
   		\begin{tabular}{lcccccccc}
       		\hline
 $\omega$ ($\mu$m) \hspace{0.5em} & d ($\mu$m) \hspace{0.5em} & L (mm) \hspace{0.5em} & R$^{\prime}$ \hspace{0.5em} & 4$\sigma^{\prime}L$ ($\mu$m) \hspace{0.5em}  &  $\sigma$/d  \hspace{0.5em} &  L$\sigma^{\prime}/r_d$ \hspace{0.5em}   &  $\varepsilon_n / \gamma \sigma$  (mrad)\hspace{0.5em}  &  $\omega /4L$ (mrad)  \\   
		\hline
       		\hline
		250	&	400	&	150	&	2.93	&	318	&	1.33	&	7.95	&	0.53	& 	27.34	\\
		150	&	200	&	100	&	2.43	&	296	&	1.92	&	7.40	&	0.71	& 	16.41	\\
		100	&	200	&	60	&	1.07	&	240	&	2.12	&	6.00	&	0.59	& 	10.94	\\
		250	&	500	&	150	&	2.41	&	840	&	1.23	&	21.00&	0.45	& 	27.34	\\
       		\hline		
   		\end{tabular}
   	}
\label{tbl:mask_design2}
\end{table*}

Among all the geometries considered, there found to be a particular geometry works for a certain envelope rather than a single design which can measure under a large range of focusing conditions. 

Examples of the resulting distributions on the screen are presented in Fig. \ref{fig:tracking}-(a), (b) and (c). The corresponding projections are shown in Fig. \ref{fig:tracking}-(d), (e) and (f) with a Gaussian curve fit to each beam-let in order to determine their mean position and widths. Emittance values are calculated using these moments and relative intensities of the beam-lets. The phase space distributions are reconstructed using divergences calculated from beam-let position offsets from the hole positions and intensity distributions of the beam-lets. These are presented in Fig. \ref{fig:tracking}-(g), (h) and (i).    

The analytical criteria were reapplied for these optimised geometries as summarised on Table \ref{tbl:mask_design2}. Even if hole diameters as small as 100$\,\mu$m considered in the initial geometries it was seen that the transmission was only a few percent proving the analysis difficult, therefore final solutions converged towards larger hole diameters. The choice of larger holes led to increase in $R^{\prime}$ values. This can be compensated by a trade off between $d$ and $L$. The choice of increasing the centre-to-centre distance between the beam-lets is limited by the number of beam-lets generated, especially in the focused distributions, where the beam size is smaller. On the other hand, the distance between the mask and the screen should be chosen carefully to minimise the overlapping. 
 
\section{Conclusions}
Apart from analytical guidelines, pepper-pot set-up requires extensive iterative numerical optimisation. A MATLAB script is developed in order to track the particle distributions, retrieved from PARMELA, through a mask with given geometry and up to a downstream screen. The script then uses the projections of the distributions at the plane, where the screen is located, to calculate the transverse emittance and reconstruct the phase space. 

During the tests on beam distributions under different solenoid fields, it was observed that for under-focused (and potentially over-focused) envelopes, pepper-pot algorithm is unable to produce retrieved results from PARMELA. This is due to the contribution from the tails of the projections when the reconstruction algorithm is used for non-Gaussian beams. This is the case when the beam emittance is not compensated, allowing emittance growth due to space charge and associated halo formation which then creates high intensity tails in the projection of the transverse plane. Therefore, the method was found to be effective for Gaussian beams under solenoid focusing close to the conditions allowing emittance compensation or matching to the linac. Furthermore, a design allowing for more than two beam-lets (a restriction especially at the focal point) is essential for statistical significance.

After demonstrating the sensitivity of pepper-pot measurement system design to incoming beam parameters, for versatile operation, we suggest the concept of a multi-region mask housing different geometries for different operational conditions.

As a merit of performance for pepper-pot technique, an expression for the systematic error is presented using the rms emittance formula and terms resulting from the pepper-pot algorithm. Once this value is smaller than the statistical deviations in the measurements, the mask design is deemed reliable for measurement precision within the given beam stability.
\vspace{1em}
\section{Acknowledgment}
This work was supported by the Cockcroft Institute Core Grant and the STFC under the project reference number ST/N00163X/1. Authors would like to thank to A. Bechold and A. Bien (NTG); D. Potkins, M. Dehnel and M. Melanson (D-PACE) and A. Palmer and A. Woolten (STFC-UKRI) for discussions on pepper-pot mask machining.

\end{document}